\documentclass[prb,showpacs,superscriptaddress]{revtex4}

\usepackage{graphicx}
\usepackage{amsmath}
\usepackage{amssymb}
\usepackage{color}
\usepackage{bm}

\begin{document}

\title{Edge states on graphene ribbon in magnetic field: interplay between
Dirac and ferromagnetic-like gaps}

\author{V. P. Gusynin}
\affiliation{Bogolyubov Institute for Theoretical Physics, 03680, Kiev, Ukraine}

\author{V. A. Miransky}
\altaffiliation[On leave from ]{Bogolyubov Institute for Theoretical Physics,
03680, Kiev, Ukraine}
\affiliation{Department of Applied Mathematics, University of Western Ontario,
London, Ontario, Canada N6A 5B7}

\author{S. G. Sharapov}
\affiliation{Bogolyubov Institute for Theoretical Physics, 03680, Kiev, Ukraine}

\author{I. A.~Shovkovy}
\affiliation{School of Applied Arts and Sciences, Arizona State University, Mesa, Arizona 85212, USA}

\author{C. M. Wyenberg}
\affiliation{Department of Applied Mathematics, University of Western Ontario,
London, Ontario, Canada N6A 5B7}

\begin{abstract}
By combining analytic and numerical methods,
edge states on a finite width graphene ribbon in a magnetic field are
studied in the framework of low-energy effective theory that takes into
account the possibility of quantum Hall ferromagnetism (QHF) gaps and
dynamically generated Dirac-like masses. The analysis is done for graphene
ribbons with both zigzag and armchair edges.
The characteristic features of
the spectrum of the edge states in both these cases are described.
In particular, the conditions for the existence
of the gapless edge states are established. Implications
of these results for the interpretation of recent experiments are discussed.
\end{abstract}

\date{\today}


\maketitle

\section{Introduction}

Graphene is a remarkable system with many unusual properties that
was created for the first time only a few years ago.\cite{Geim2004Science}
(For reviews on graphene see, for example, Refs.~\onlinecite{review1,review2,
review3,review4}.) One of such properties is an unconventional quantum
Hall effect (QHE).
{Instead of the ordinary QHE, expected in the case
of 2-dimensional non-relativistic electron systems, an anomalous
quantization is observed in graphene.}\cite{Geim2005Nature,Kim2005Nature}
The observation appears to be in perfect agreement with the theoretical
predictions,\cite{Ando2002,Gusynin2005PRL,Peres2005} stating that
the QHE plateaus should occur at filling factors $\nu = \pm 4(|n|+1/2)$
where $n$ is an integer. This anomalous  quantization is a direct
outcome of the relativistic-like nature of the low-energy quasiparticles
in graphene described by a Dirac theory with an internal $U(4)$
symmetry.\cite{Semenoff1984PRL,Haldane1988PRL}

It is the $U(4)$ symmetry, operating in the spin and sublattice-valley
spaces, that is responsible for the 4-fold degeneracy of the Landau
energy levels and for the overall factor 4 in the filling factors of the
observed plateaus, $\nu = \pm 4(|n|+1/2)$. Strictly speaking, the
$U(4)$ symmetry is not exact but broken down to a smaller
$U(2)_{+} \times U(2)_{-}$ symmetry group by the Zeeman term.
The latter symmetry operates in the sublattice-valley space and
does not mix spin-up ($s=+$) and spin-down ($s=-$) states.
When the magnetic field is not too strong, a relatively small Zeeman
term does not affect the observable QHE in a qualitative way.

When the magnetic field becomes sufficiently strong, the QHE plateaus
$\nu = 0, \pm 1, \pm 4$ are observed.\cite{ZhangPRL,JiangPRL,AbaninNovoselov,Ong2007}
This suggests that the
4-fold degeneracy of the Landau levels is lifted.
The new plateaus may
be explained by one of the following seemingly different
theoretical scenarios. (i) Quantum Hall ferromagnetism (QHF)
\cite{qhf1,qhf2,qhf3,qhf4,qhf5}, which is
connected with the theory of exchange-driven spin splitting of Landau
levels.{\cite{Fogler1995}} The QHF
order parameters are densities of the conserved charges connected
with three diagonal generators of the non-abelian subgroup $SU(4)\subset U(4)$
(the dynamics of a Zeeman spin splitting
enhancement considered in Ref.~\onlinecite{Abanin} is intimately connected with
the QHF). (ii) The magnetic catalysis (MC) scenario
\cite{dm1,dm2,dm3,dm4} that is
based on the phenomenon of an enhancement of the density of states in infrared
by a magnetic field, which catalyzes
electron-hole pairing (leading to excitonic condensates) in relativistic-like
systems.\cite{magneticcatalysis,mc1,mc2}. This scenario
invokes electron-hole
pairing and excitonic condensates to produce dynamically generated Dirac-like
masses in the low-energy theory.

Recently, by analyzing a gap equation for the propagator of Dirac
quasiparticles, it has been found in Refs.~\onlinecite{GGM2007,GGMS2008}
that the QHF and MC order parameters {\it necessarily} coexist. As
will be shown in the present paper, this feature could have important
consequences for the dynamics of edge states in the QHE in graphene.

The study of edge states is of general interest because such
states provide a deeper insight into the quantum Hall effect.\cite{Halperin1982}
Currently there exist many studies of edge
states in graphene-like
systems.\cite{Peres2005,Nakada,Fujita,Falko,KaneMele,Brey,Brey1,PCNG2006,Abanin,Peeters,GMSS2008,Arikawa}
In Refs.~\onlinecite{Abanin,Brey},
it was found that in the presence of a
magnetic field there may exist
gapless modes of such states and they should play
an important role in charge transport of graphene near the Dirac
neutral point. The gapless modes were shown to appear when the
lowest Landau level (LLL) was split by a spin gap, or in other
words, by enhanced Zeeman splitting leading to the $\nu = 0$ plateau.
Such gapless states were
absent, though, in the case of a Dirac mass gap.\cite{AbaninNovoselov,Abanin}

While the presence of the gapless edge states should make graphene a
so-called quantum Hall metal, their absence should make it an
insulator.\cite{AbaninNovoselov,Abanin} The actual temperature dependence of
the longitudinal resistivity at the $\nu=0$ plateau in Refs.~\onlinecite{ZhangPRL,AbaninNovoselov}
is consistent with the metal type. Thus, it was argued that the origin of
the $\nu=0$ plateau is connected with the enhanced spin (ferromagnetic)
splitting of the LLL.\cite{AbaninNovoselov,Abanin}

The conclusion of Ref.~\onlinecite{AbaninNovoselov,Abanin} regarding the
origin of the $\nu=0$ plateau does not appear to be universal
however. The recent data from Ref.~\onlinecite{Ong2007} reveal a clear
plateau at $\nu=0$, but the temperature dependence of the diagonal
component of the resistivity signals a crossover to an insulating
state in high fields. This does not seem to support the existence
of gapless edge states. So, one may ask whether there is indeed a
Dirac-type mass gap and no spin gap in the device studied in
Ref.~\onlinecite{Ong2007}?

Motivated by this question and a theoretical analysis in Refs.~\onlinecite{GGM2007,GGMS2008},
the spectrum of edge states {has been recently}
studied under the assumption that the removal of sublattice
and spin degeneracies in graphene in a strong magnetic field is
connected with the generation of
{\it both} Dirac masses and spin gaps.\cite{GMSS2008} The main result
of that work was establishing a criterion for the existence of the gapless
edge states in the QHE in graphene on a half-plane with a zigzag or
armchair edge.
It was concluded that the controversy between the experimental results
in Refs.~\onlinecite{AbaninNovoselov} and \onlinecite{Ong2007} may reflect more rich and
complicated dynamics in the QHE in graphene than those considered in the
QHF and MC scenarios.

In the present work, we extend the analysis in Ref.~\onlinecite{GMSS2008} to
the case of a finite width graphene ribbon.
The characteristic feature
in our approach is combining analytic and numerical methods in the analysis
of the edge states. This allows to describe the main features of the
quasiparticle spectrum and, in particular, to extend the
criterion obtained in Ref.~\onlinecite{GMSS2008} to a more complicated and
interesting case of a graphene sample with two boundaries, which in turn
yields a deeper insight in physics behind this criterion.

The paper is organized as follows. In {Sec.~\ref{secII}}, we discuss the low-energy
field-theoretical model of graphene with MC and QHF dynamical order parameters used
in the rest of the paper. An overview of the general formalism for studying edge
states of a finite width graphene ribbon in an external magnetic field is given
in Sec.~\ref{secIII}. The numerical analysis in the case of zigzag and armchair
edges is presented in Sec.~\ref{zigzag-num} and \ref{armchair-num}, respectively.
The discussion of the main results and their experimental implications are given
in Sec.~\ref{Discussion}.

\section{QHF and MC order parameters in graphene}
\label{secII}

{
For convenience of the analysis of the edge state, the chiral
representation of the Dirac matrices will be used here (see, for
example, Ref.~\onlinecite{review4}),
\begin{eqnarray}
\gamma^0&=&\tilde{\tau}^{1}\otimes\tau^{0}=\left(
\begin{array}{cc} 0 & I \\
                            I & 0\end{array}\right),\quad
\gamma^i=-i\tilde{\tau}^{2}\otimes \tau^{i}=\left(
\begin{array}{cc} 0 & -\tau^i \\
                             \tau^i & 0\end{array}\right),\\
\gamma_5&\equiv&i\gamma^0\gamma^1\gamma^2\gamma^3=\tilde{\tau}^{3}\otimes\tau^{0}
=\left(
\begin{array}{cc} I & 0 \\
                            0 & -I\end{array}\right),
\label{chiral_repres}
\end{eqnarray}
where $\tilde{\tau}^{i},\tau^i$ are Pauli matrices and $\tau^{0}$ is the $2\times2$
unit matrix.}
The spinor field of Dirac quasiparticles ${\Psi^{s}}^T =
\left(\psi_{K_{+}A}^s, \psi_{K_{+}B}^s, \psi_{K_{-}B}^s,
\psi_{K_{-}A}^s\right)$
combines the Bloch states on two valleys ($K_{+}$ and $K_{-}$) and
two sublattices ($A$ and $B$), $s$ is the spin index.
{The QHF order parameters are the spin density
$\langle\Psi^{\dagger}P_s\Psi\rangle$ and the pseudospin density
$\langle\Psi^{\dagger}\gamma^5P_s\Psi\rangle$,
with
$P_{\pm}=(1 \pm \sigma^3)/2$ being projectors on states with spin directed
along (+) and opposite ($-$) the
magnetic field. These order parameters are related to the
chemical potentials $\mu_s$ and
$\tilde{\mu}_s$, respectively. On the other hand, the MC order
parameter is the vacuum expectation value of the Dirac mass term
$\langle\bar{\Psi}\gamma^3P_s\Psi\rangle$ associated with the
conventional Dirac mass $\tilde{\Delta}_s$ (here $\bar\Psi=\Psi^{\dagger}\gamma^{0}$).

Recently, a unifying approach, combining and augmenting both
QHF and MC mechanisms, was proposed in Refs.~\onlinecite{GGM2007,GGMS2008}.
By analyzing the gap equation with a local Coulomb interaction and
using a multi-parameter variational ansatz
for the quasiparticle propagator, it was found that {(i)} the new
MC order parameter $\langle\bar{\Psi}\gamma^3\gamma^5P_s\Psi\rangle$,
related to a Dirac mass $\Delta_s$ that breaks time reversal
symmetry \cite{Haldane1988PRL}, has to be added, and {(ii)} the QHF and MC
order parameters necessarily coexist. (Let us emphasize
that, in the presence of an external magnetic field, the time-reversal
symmetry is broken and a state with the vanishing $\Delta_s$ is not
protected by any symmetry.)}

{More precisely, it was shown in Refs.~\onlinecite{GGM2007,GGMS2008} that 
for a fixed spin, the full inverse quasiparticle propagator takes the following 
general form (in the chiral representation used in this paper):
\begin{eqnarray}
iG^{-1}_{s}(u,u^\prime)&=& \left[(i\hbar\partial_t+\mu_{s} +
\tilde{\mu}_{s}\gamma^5)\gamma^0 - v_{F}(\bm{\pi}\cdot\bm{\gamma})
-\tilde{\Delta}_{s}\gamma^3 + \Delta_{s}\gamma^3\gamma^5\right]\delta^{3}(u-u^\prime),
\label{full-inverse}
\end{eqnarray}
where $\bm{\pi}$ is the canonical momentum and the parameters $\mu_{s}$, 
$\tilde{\mu}_{s}$, $\Delta_{s}$, and $\tilde{\Delta}_{s}$ are determined from 
the gap equation. Note that the {full} electron chemical potentials $\mu_{\pm}$ 
include the Zeeman energy $\mp Z$ with 
\begin{equation}
Z \simeq \mu_BB = 0.67B[\mbox{T}]~\mbox{K}.
\label{Zeeman}
\end{equation}}

By making use of the explicit form of the spinor, we find the following
correspondence between the four types of order parameters and the
electron densities (the spin index is omitted):
\begin{eqnarray}
\mu &\to&\langle{\Psi}^{\dagger}\Psi\rangle = n_{K_{+}A}+n_{K_{-}A}+
n_{K_{+}B}+n_{K_{-}B},\\
\tilde\mu &\to &\langle{\Psi}^{\dagger}\gamma^{5}\Psi\rangle =
n_{K_{+}A}-n_{K_{-}A}+n_{K_{+}B}-n_{K_{-}B},\\
\Delta &\to &\langle\bar{\Psi}\gamma^{3}\gamma^{5}\Psi\rangle =
n_{K_{+}A}-n_{K_{-}A}-n_{K_{+}B}+n_{K_{-}B},\\
\tilde\Delta &\to &\langle\bar{\Psi}\gamma^{3}\Psi\rangle =
 n_{K_{+}A}+n_{K_{-}A}-n_{K_{+}B}-n_{K_{-}B},
\end{eqnarray}
where the $n_{K_{+}A}$, $n_{K_{-}A}$, $n_{K_{+}B}$ and $n_{K_{-}B}$ are 
the densities of quasiparticles at specified valleys and sublattices.
The QHF order parameters associated with $\mu $ and $\tilde\mu $ are
the total density of electrons with a given spin and the density imbalance
between the two valleys, respectively. Note that the MC order parameter 
related to the conventional Dirac mass $\tilde\Delta$ describes the 
density imbalance between the $A$ and $B$ sublattices (i.e., a charge
density wave as interpreted in Refs.~\onlinecite{mc1,dm1,dm2,dm3,dm4}).
{The value of the singlet Dirac mass $\Delta$ [see Eq.~(7)] controls
a mixed density imbalance at the two valleys and the two sublattices.}

In terms of symmetry, these order parameters can be divided in
two groups. The order parameters $\langle\Psi^{\dagger}P_s\Psi\rangle$
and $\langle\bar{\Psi}\gamma^3\gamma^5P_s\Psi\rangle$
(related to $\mu_s$ and $\Delta_s$) with nonequal values for $s=\pm$
break the approximate $U(4)$ symmetry just like the Zeeman term, and
they are singlets under the non-abelian subgroups $SU(2)_s\subset U(2)_{s}$
{with the generators $P_{s}\otimes ( -i\gamma^{3}/2,\gamma^{3}\gamma^{5}/2,
\gamma^{5}/2)$ }.
Since these singlet order parameters break no exact symmetries of the
action, they are not the order parameters in the strict sense. Yet, because
of a relative smallness of the ``bare'' Zeeman energy $Z$ and a significant
dynamical part in $\mu_s$ and $\Delta_s$, it is appropriate to talk about
approximate spontaneous symmetry breaking. In the model in
Refs.~\onlinecite{GGM2007,GGMS2008}, these two order parameters coexist and
play a crucial role in the solution corresponding to the $\nu = 0$ plateau.

The order parameters of the other type,
$\langle\Psi^{\dagger}\gamma^5P_s\Psi\rangle$
and $\langle\bar{\Psi}\gamma^3P_s\Psi\rangle$,
(associated with $\tilde\mu_s$ and $\tilde\Delta_s$)
are triplets under $SU(2)_s\subset U(2)_{s}$. Each of these two order
parameters describes spontaneous $SU(2)_{s}$ symmetry breaking down to
$U(1)_{s}$ with the generator $P_{s}\otimes \gamma^{5}/2$. 
In the model in Refs.~\onlinecite{GGM2007,GGMS2008}, these two
order parameters coexist and play an important role in the solution
corresponding to the $\nu =\pm 1$ plateaus
{(as well as to the plateaus $\nu =\pm 3$
and $\nu =\pm 5$, which have not been observed yet).}

\section{Landau levels and edge states}
\label{secIII}

In accordance with the discussion in the previous section,
we assume that both ferromagnetic and mass type gaps
coexist in general. Our goal is to find the spectrum of edge
states in such a theory.

{The structure of the inverse quasiparticle propagator in 
Eq.~(\ref{full-inverse}) implies that in the most general case
the quadratic part of the effective Hamiltonian for quasiparticles
of spin $s$ takes the following form in the first quantized theory:}
\begin{equation}
\label{Hamiltonian} \hat{H}_s = \hat{H}_0 -\mu_s -\tilde\mu_s
\gamma^5 -\Delta_s\gamma^0\gamma^3\gamma^5 +\tilde\Delta_s
\gamma^0\gamma^3.
\end{equation}
The free part of the Hamiltonian reads
\begin{equation}
\hat{H}_0= v_F \left(\alpha_1 \pi_x+\alpha_2 \pi_y\right)
\end{equation}
where $v_F\simeq 10^6~\mbox{m/s}$ is the Fermi velocity. By definition,
$\alpha_i= \gamma^0\gamma^i$ and the canonical momentum is
$\pi_{i}=-i\hbar \partial_i+eA_i/c$. Here the vector potential is
taken in the Landau gauge: $A_x=-By$ and $A_y=0$, where $B$ is the
magnitude of a constant magnetic field orthogonal to the $xy$ plane of
graphene.

The parameters $\mu_{s}$, $\tilde{\mu}_{s}$, $\Delta_{s}$ and $\tilde{\Delta}_{s}$ are
determined from the gap equation. In particular, in the model in
Refs.~\onlinecite{GGM2007,GGMS2008},
{the $S1$ solution (in the nomenclature of
Ref.~\onlinecite{GGMS2008})} near the Dirac neutral point, corresponding
to the $\nu =0$ plateau, has the following form:
\begin{equation}
\tilde{\Delta}_\pm= \tilde{\mu}_\pm=0,\,\, \mu_\pm  = \mp Z \mp
A,\,\, \Delta_\pm = \pm M\, \label{singlet},
\end{equation}
where for the values of magnetic fields $B \lesssim 45\, \mbox{T}$ utilized in
the experiments in Refs.~\onlinecite{ZhangPRL,JiangPRL,AbaninNovoselov,Ong2007},
the dynamical parameters $A$, $M$ are considerably larger than $Z$, and $M > A$.
The value of the spin gap in this solution is $\Delta E = 2M + 2(Z +A)$.
Note that it is essentially larger than the spin gap $\Delta E = 2(A + Z)$
in the QHF scenario.\cite{Abanin} While the problem of calculating the values
of $\mu_{s}$, $\tilde{\mu}_{s}$, $\Delta_{s}$ and $\tilde{\Delta}_{s}$ is
not addressed in this study, it should be clear that in an actual
device they are determined by (i) the strength of the magnetic
field, (ii) the temperature, and (iii) other sample-specific
parameters (e.g., the mobility of carriers, the size and geometry,
the type of the substrate etc.). In practice, we analyze the
spectrum of edge states in the model described by the model
Hamiltonian in Eq.~(\ref{Hamiltonian}).

{In general, in a finite geometry case, the magnitude of the 
exchange and Hartree interactions which determine the values of the 
parameters $\Delta_s$, $\tilde{\Delta}_s$, $\mu_s$, and $\tilde{\mu}_s$ 
is likely to vary with the distance from the edge and should be calculated 
in a self-consistent way. The present study of the edge states is done 
assuming uniform gaps and uniform chemical potentials, and so it captures 
qualitative, but probably not quantitative aspects of the edge-state 
physics (see also a related discussion in the end of Sec.~\ref{Discussion})}.

When written in components, the Dirac equation corresponding
to the Hamiltonian (\ref{Hamiltonian}) takes the
following form:
\begin{eqnarray}
\left(\begin{array}{@{\extracolsep{-1mm}}cc}
    E+\mu^{(+)}+\Delta^{(-)} &  \hbar v_{F}\left(iD_{x}+D_{y}\right) \\
    \hbar v_{F}\left(iD_{x}-D_{y}\right) &E+\mu^{(+)}-\Delta^{(-)}
\end{array}\hspace{-2mm}\right)
\left(\begin{array}{@{\extracolsep{-1mm}}c}\psi_{K_{+}A}\\
\psi_{K_{+}B}\end{array}\hspace{-2mm}\right) = 0,
\label{eq-masses:Kplus}\\
\left( \begin{array}{@{\extracolsep{-1mm}}cc}
    E+\mu^{(-)}+\Delta^{(+)} &  -\hbar v_{F}\left(iD_{x}+D_{y}\right) \\
    -\hbar v_{F}\left(iD_{x}-D_{y}\right) &E+\mu^{(-)}-\Delta^{(+)}
\end{array}\hspace{-2mm}\right)
\left(\begin{array}{@{\extracolsep{-1mm}}c}\psi_{K_{-} B}\\
\psi_{K_{-} A}\end{array}\hspace{-2mm}\right)
= 0.
\label{eq-masses:Kminus}
\end{eqnarray}
Here {the covariant derivative $D_i = \partial_i+ (ie/{\hbar c})A_i$} and
the shorthand notations $\mu^{(\pm)}\equiv \mu\pm\tilde\mu$
and $\Delta^{(\pm)}\equiv \Delta\pm\tilde\Delta$ were introduced (the spin
index $s$ was omitted). In each of the two sets of equations, the $B$-components
of the wave function can be eliminated,
\begin{eqnarray}
\psi_{K_{+}B} &=&  \frac{\hbar v_{F}\left(-iD_{x}+D_{y}\right)\psi_{K_{+}A}}{E+\mu^{(+)}-\Delta^{(-)}}  ,\\
 \psi_{K_{-} B}&=&\frac{\hbar v_{F}\left(iD_{x}+D_{y}\right) \psi_{K_{-} A}}{E+\mu^{(-)}+\Delta^{(+)}}.
\end{eqnarray}
By taking these into account, we derive the equations for the $A$-components,
\begin{eqnarray}
\left(-l^{2}D_{x}^{2}-l^{2}D_{y}^{2}+1\right)\psi_{K_{+}A}&=&2\lambda_{+}\psi_{K_{+}A} ,\\
\left(-l^{2}D_{x}^{2}-l^{2}D_{y}^{2} -1\right)\psi_{K_{-} A} &=&2\lambda_{-}\psi_{K_{-} A},
\end{eqnarray}
where $\lambda_{\pm}=\left[\left(E+\mu^{(\pm)}\right)^{2}
-\left(\Delta^{(\mp)}\right)^{2}\right]/\epsilon_0^{2}$,
$l\equiv \sqrt{\hbar c /|eB|}$ is the magnetic length, and
$\epsilon_0\equiv \sqrt{2\hbar v_F^2 |eB|/c}$ is the Landau
energy scale. Note that $\lambda_{+}$ ($\lambda_{-}$) is related to
the $K_{+}$ ($K_{-}$) valley.

In the Landau gauge $\mathbf{A}=(-By,0)$, the wave functions are plane waves in
the $x$-direction. 
Thus, we write
\begin{eqnarray}
\psi_{K_{+}A}(\mathbf{r},k) =  \frac{e^{ikx}}{\sqrt{2\pi l}} u_{+}(y,k), \quad
\psi_{K_{+}B}(\mathbf{r},k) = \frac{e^{ikx}}{\sqrt{2\pi l}} v_{+}(y,k),
&&
\\
 \psi_{K_{-} A}(\mathbf{r},k) = \frac{e^{ikx}}{\sqrt{2\pi l}} u_{-}(y,k), \quad
  \psi_{K_{-} B}(\mathbf{r},k) =\frac{e^{ikx}}{\sqrt{2\pi l}} v_{-}(y,k).
&&
\end{eqnarray}
The envelope functions $u_{\pm}(y,k)$ and $v_{\pm}(y,k)$ depend
only on a single combination of the variables, $\xi=y/l -kl$, and
satisfy the following equations:
\begin{eqnarray}
\label{eq.u}
&&
\left(\partial^{2}_{\xi}-\xi^{2}\mp1+2\lambda_{\pm} \right)u_{\pm}(\xi)=0,\\
\label{eq.v}
&&
v_{\pm}(\xi) =\frac{\epsilon_0\left(\partial_{\xi}\mp\xi\right)u_{\pm}(\xi)}
{\sqrt{2}\left( E+\mu^{(\pm)}\mp\Delta^{(\mp)}\right)}.
\end{eqnarray}
Note that the wavevector $k$ determines the center of the electron
orbital along the $y$-direction, $y_k=kl^2$. Then, as we shall see
below, for a system with a ribbon geometry, e.g., $0 \leq y\leq
W$, the condition of finite energy will be satisfied only for
eigenstates with wavevectors $k$ in a finite range, $0\lesssim
k\lesssim W/l^2$. This is known as the position-wavevector duality
in the Landau gauge. (Note that the {maximum} value of the wavevector
$k$, measuring the displacement from either $K_{+}$ or $K_{-}$
point, is limited by the boundaries of the first Brillouin zone.
However, this fact is not explicit in the low-energy theory.)

The general solution to Eqs.~(\ref{eq.u}) and (\ref{eq.v}) is given in
terms of the parabolic cylinder (Weber) functions $U(a,z)$ and $V(a,z)$,\cite{Abramowitz}
\begin{eqnarray}
\label{uPlus}
&&\hspace{-12mm}
u_{+}(\xi)= C_{1} \frac{E+\mu^{(+)}-\Delta^{(-)}}{\epsilon_0 }
U\left(\frac{1-2\lambda_{+}}{2},\sqrt{2}{\xi}\right)
+C_{2} V\left(\frac{1-2\lambda_{+}}{2},\sqrt{2}{\xi}\right),\\
\label{vPlus}
&&\hspace{-12mm}
v_{+}({\xi})=-C_{1}
U\left(-\frac{1+2\lambda_{+}}{2},\sqrt{2}{\xi}\right)
-C_{2}\frac{E+\mu^{(+)}+\Delta^{(-)}}{\epsilon_0 }
V\left(-\frac{1+2\lambda_{+}}{2},\sqrt{2}{\xi}\right),\\
\label{uMinus}
&&\hspace{-12mm}
u_{-}(\xi) = C_{3}U\left(-\frac{1+2\lambda_{-}}{2},\sqrt{2}{\xi}\right)
+C_{4} \frac{E+\mu^{(-)}+\Delta^{(+)}}{\epsilon_0}
V\left(-\frac{1+2\lambda_{-}}{2},\sqrt{2}{\xi}\right),\\
\label{vMinus}
&&\hspace{-12mm}
v_{-}({\xi})=C_{3} \frac{E+\mu^{(-)}-\Delta^{(+)}}{\epsilon_0}
U\left(\frac{1-2\lambda_{-}}{2},\sqrt{2}{\xi}\right)
+C_{4}V\left(\frac{1-2\lambda_{-}}{2},\sqrt{2}{\xi}\right).
\end{eqnarray}
Note the following relations with the parabolic cylinder
functions $D_{\nu}(z)$:
\begin{eqnarray}
U(a,z) &=& D_{-a-1/2}(z),\\
V(a,z) &=& \frac{\Gamma(a+1/2)}{\pi}\left[
\sin(\pi a)D_{-a-1/2}(z)+D_{-a-1/2}(-z)
\right].
\end{eqnarray}
In an infinite system, the normalizable wave functions contain
only the parabolic cylinder $U$-functions which are bound at
$z\to\pm\infty$ provided $a=-n-1/2$ and $n$ is a non-negative
integer. In fact, the following relation is valid:
$U(-n-1/2,z)=2^{-n/2} e^{-z^2/4}H_n(z/\sqrt{2})$,
where $H_n(z)$ are the Hermite polynomials. In this case, the spectrum
is determined by $\lambda^{(bulk)}_{\pm}=n$ with
$n=0,1,2,\ldots$.

\begin{figure}
\includegraphics[width=.48\textwidth]{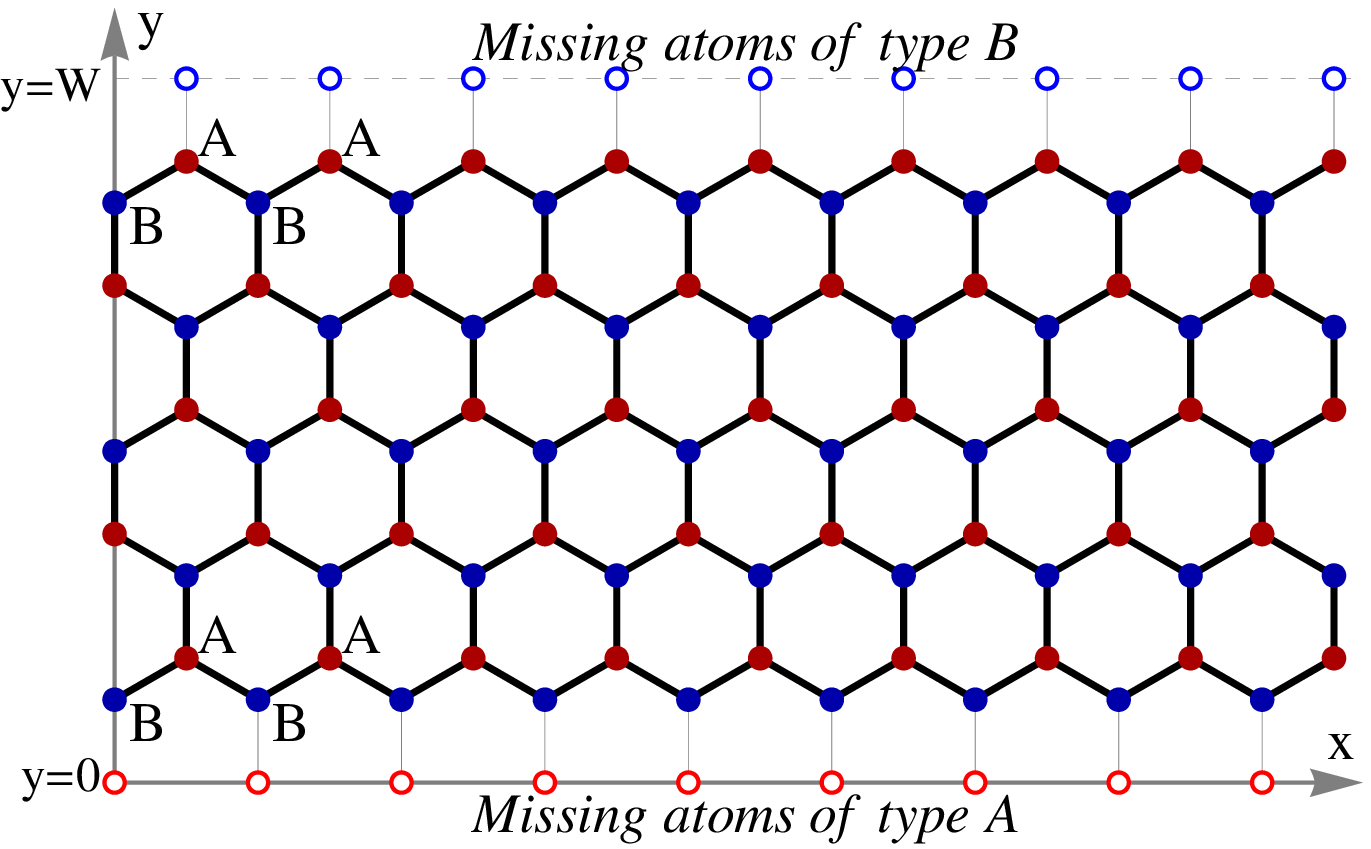}
\hspace{.05\textwidth}
\includegraphics[width=.45\textwidth]{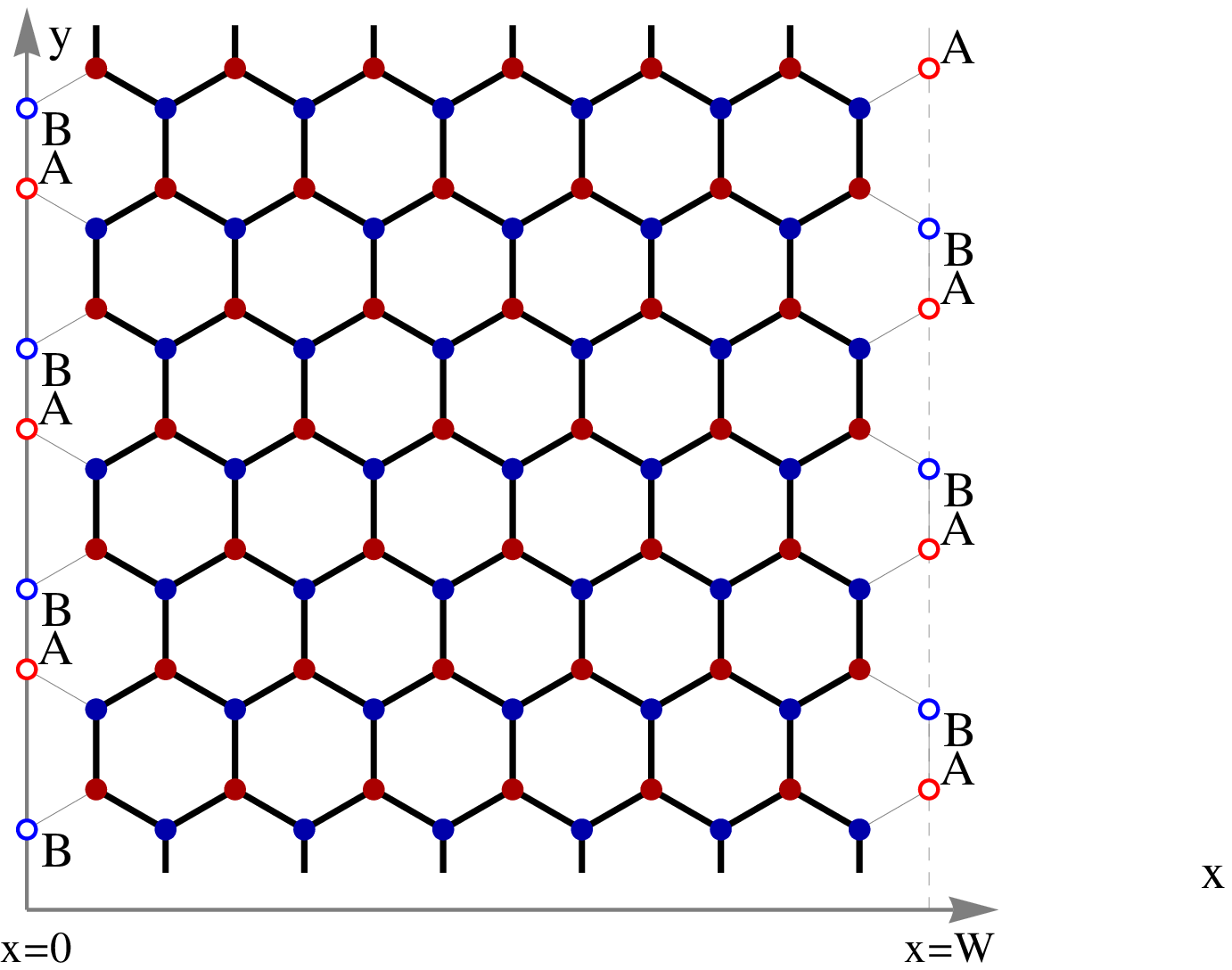}
\caption{(Color online) The lattice structure of a finite width graphene ribbon with
zigzag (left panel) and armchair (right panel) edges.}
\label{fig.1}
\end{figure}

{In the case of a graphene ribbon of a finite width in the $y$-direction,
$0 \leq y\leq W$, and with two {\it zigzag} edges parallel to the $x$-direction,}
the $A$- and $B$-components of wave functions
should vanish on the opposite edges \cite{Brey,Abanin}, i.e.,
\begin{eqnarray}
\label{boundary0}
&&\hspace{-12mm}
y=0:\quad u_{+}(-kl)=u_{-}(-kl) = 0 , \\
\label{boundaryW}
&&\hspace{-12mm}
y=W:\quad  v_{+}(W/l-kl)=v_{-}(W/l-kl) = 0,
\end{eqnarray}
see Fig.~\ref{fig.1}
In principle, by satisfying these equations and using the wave function
normalization conditions, we can determine all four integration constants
in Eqs.~(\ref{uPlus})--(\ref{vMinus}). For our purposes here, however, it
suffices to determine the conditions when non-trivial, normalizable
solutions exist. These will provide the dispersion spectra of all modes
in a ribbon of graphene. The corresponding numerical analysis is presented
in Sec.~\ref{zigzag-num}.

In the case of a graphene ribbon with {\it armchair} edges parallel to
the $y$-direction, it is convenient to choose a different Landau gauge
with $(A_{x},A_{y})=(0,Bx)$. Accordingly, the solutions are translation
invariant along the $y$-direction,
\begin{eqnarray}
\label{y-armchair}
\psi_{AK} (\mathbf{r},k)&=&\frac{1}{\sqrt{2\pi
l}}\,e^{iky}u_{+}(x,k),\qquad \psi_{BK}(\mathbf{r},k)=\frac{1}{\sqrt{2\pi
l}}\,e^{iky}v_{+}(x,k), \\
\psi_{AK_{-}}(\mathbf{r},k)&=&\frac{1}{\sqrt{2\pi
l}}\,e^{iky}u_{-}(x,k), \qquad \psi_{BK_{-}}(\mathbf{r},k)=\frac{1}{\sqrt{2\pi
l}}\,e^{iky}v_{-}(x,k).
\end{eqnarray}
Then, the corresponding differential equations for functions
$u_{\pm}(x,k)$, which are rewritten in terms of the dimensionless variable
$\xi=x/l +kl$, coincide with Eq.~(\ref{eq.u}). The expressions for the
eliminated components $v_{\pm}(\xi)$, however, slightly differ from
Eq.~(\ref{eq.v}), and are given by
\begin{equation}
\label{v_pm-armchair}
v_{\pm}(\xi) =\mp
i\frac{\epsilon_0\left(\partial_{\xi}\mp\xi\right)u_{\pm}(\xi)}
{\sqrt{2}(E+\mu^{(\pm)} \mp\Delta^{(\mp)})}.
\end{equation}
The general solutions for the $u_{\pm}(\xi)$ functions have the same form
as in Eqs.~(\ref{uPlus}) and (\ref{uMinus}),
\begin{eqnarray}
\label{uPlusArm}
&& u_{+}(\xi)= C_{1} \frac{E+\mu^{(+)}-\Delta^{(-)}}{\epsilon_0 }
U\left(\frac{1-2\lambda_{+}}{2},\sqrt{2}{\xi}\right)
+C_{2} V\left(\frac{1-2\lambda_{+}}{2},\sqrt{2}{\xi}\right),\\
\label{uMinusArm}
&& u_{-}(\xi) = C_{3}U\left(-\frac{1+2\lambda_{-}}{2},\sqrt{2}{\xi}\right)
+C_{4} \frac{E+\mu^{(-)}+\Delta^{(+)}}{\epsilon_0}
V\left(-\frac{1+2\lambda_{-}}{2},\sqrt{2}{\xi}\right),
\end{eqnarray}
but with $\xi=x/l +kl$. By using the relations in Eq.~(\ref{v_pm-armchair}),
we also obtain the explicit expression for $v_{\pm}(\xi)$
functions,
\begin{eqnarray}
\label{vPlusArm}
&&
v_{+}({\xi})=iC_{1}
U\left(-\frac{1+2\lambda_{+}}{2},\sqrt{2}{\xi}\right)
+iC_{2}\frac{E+\mu^{(+)}+\Delta^{(-)}}{\epsilon_0 }
V\left(-\frac{1+2\lambda_{+}}{2},\sqrt{2}{\xi}\right),\\
\label{vMinusArm}
&&
v_{-}({\xi})=iC_{3} \frac{E+\mu^{(-)}-\Delta^{(+)}}{\epsilon_0}
U\left(\frac{1-2\lambda_{-}}{2},\sqrt{2}{\xi}\right)
+iC_{4}V\left(\frac{1-2\lambda_{-}}{2},\sqrt{2}{\xi}\right).
\end{eqnarray}
Since the armchair edges have lattice sites of both $A$ and $B$ types, the wave
function should vanish both at the $x=0$ and $x=W$
lines,\cite{Falko,Brey,Abanin}
\begin{eqnarray}
x=0:  &\quad & u_{+}(kl)+u_{-}(kl)=0,  \qquad  \qquad  \qquad  \quad v_{+}(kl)+v_{-}(kl)=0,
\label{BC-arm1}  \\
x=W:  &\quad & u_{+}(W/l+kl)+u_{-}(W/l+kl)=0, \quad v_{+}(W/l+kl)+v_{-}(W/l+kl)=0.
\label{BC-arm2}
\end{eqnarray}
It is important to notice that the armchair boundary conditions mix the chiralities
associated with the $K_{+}$ and $K_{-}$ valleys, which makes the analysis more
involved than in the case of zigzag edges. The details are given in
Sec.~\ref{armchair-num}.

\section{Numerical results in the case of zigzag edges.}
\label{zigzag-num}

\begin{figure*}
\begin{center}
\includegraphics[width=.45\textwidth]{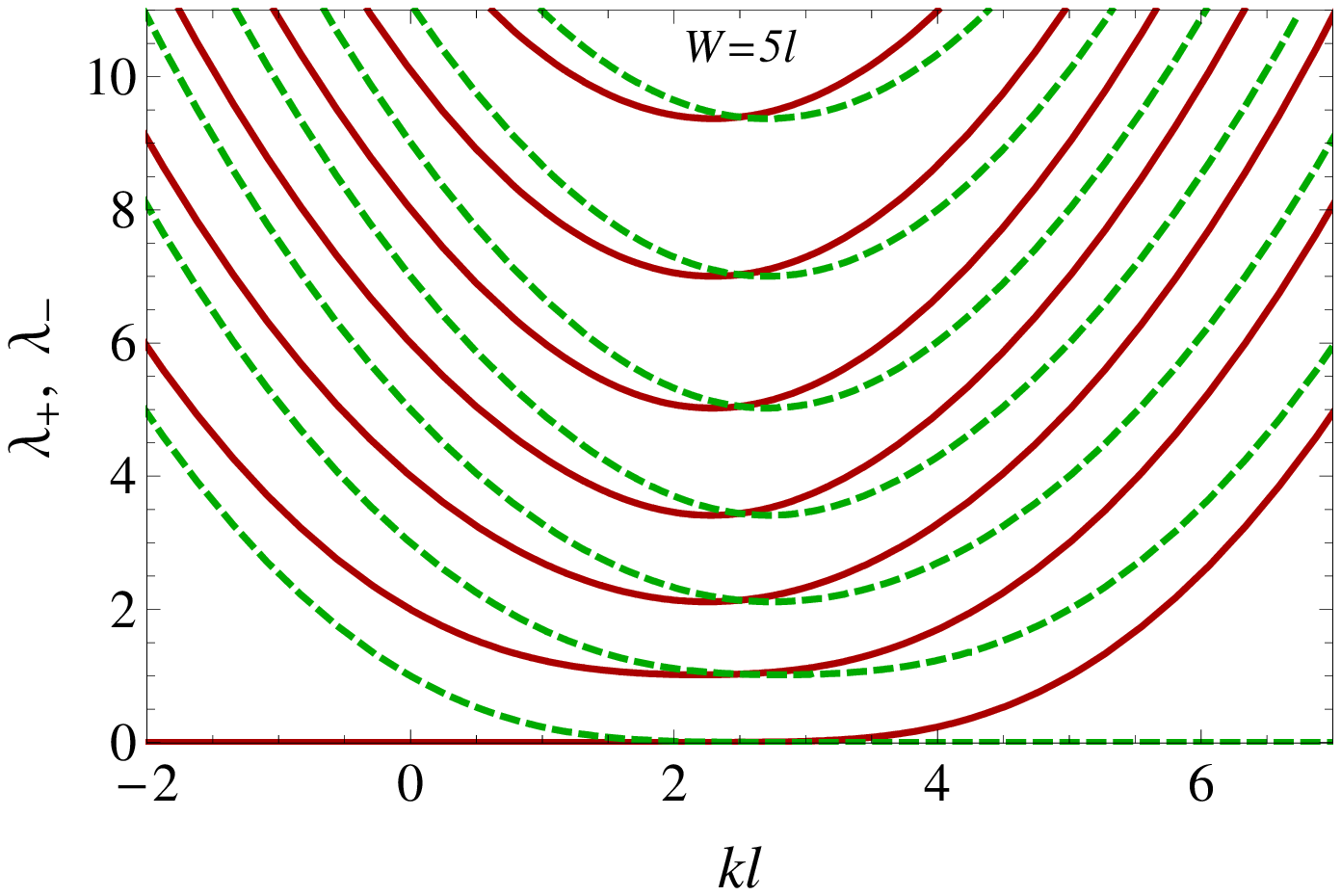}\hspace{.05\textwidth}
\includegraphics[width=.45\textwidth]{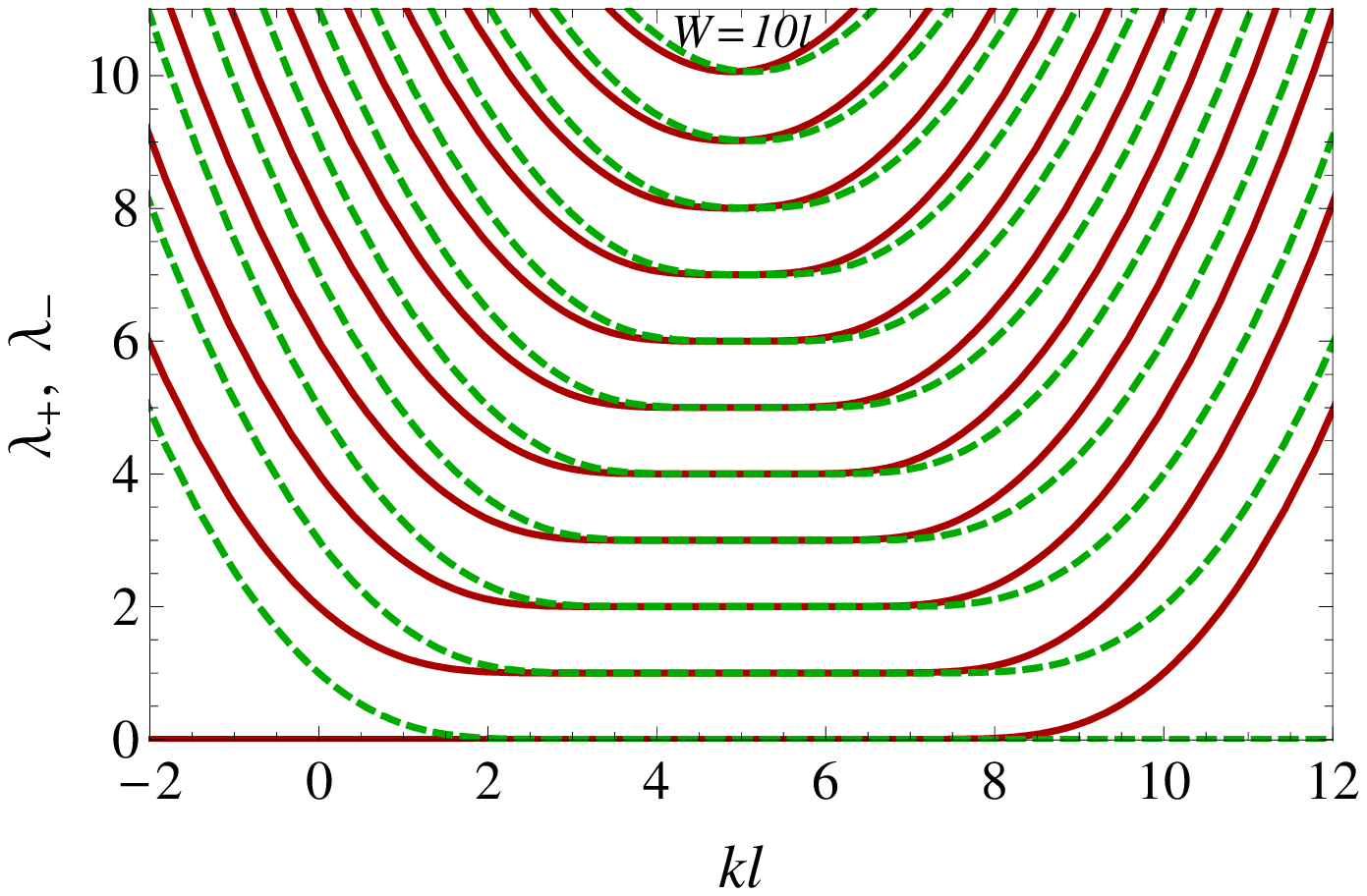}
\caption{(Color online) Numerical results for the dimensionless parameters $\lambda_{+}$ (solid
lines) and $\lambda_{-}$ (dashed lines) in the case of ribbons with zigzag edges.
The ribbons' widths are $W=5l$ (left panel) and $W=10l$ (right panel).}
\label{fig.2}
\end{center}
\end{figure*}

Let us start from the boundary conditions at $K_{+}$ valley,
see (\ref{boundary0}) and (\ref{boundaryW}). They take the
following explicit form:
\begin{eqnarray}
\label{cond1}
&&\hspace{-12mm}
C_{1}\frac{E+\mu^{(+)}-\Delta^{(-)}}{\epsilon_0 }
U\left(\frac{1-2\lambda_{+}}{2},-\sqrt{2}kl\right)  
=-C_{2}V\left(\frac{1-2\lambda_{+}}{2},-\sqrt{2}kl\right),\\
\label{cond2}
&&\hspace{-12mm}
C_{1}U\left(-\frac{1+2\lambda_{+}}{2},\sqrt{2}(k_0-k)l\right)  
=-C_{2}\frac{E+\mu^{(+)}+\Delta^{(-)}}{\epsilon_0 }
V\left(-\frac{1+2\lambda_{+}}{2},\sqrt{2}(k_0-k)l\right),
\end{eqnarray}
where $k_0\equiv W/l^2$ is determined by the width of the ribbon.
A nontrivial solution to this set of equations exists when
\begin{eqnarray}
&&\hspace{-10mm}
\lambda_{+}
U\left(\frac{1-2\lambda_{+}}{2},-\sqrt{2}kl\right)
V\left(-\frac{1+2\lambda_{+}}{2},\sqrt{2}(k_0-k)l\right)
-U\left(-\frac{1+2\lambda_{+}}{2},\sqrt{2}(k_0-k)l\right)
V\left(\frac{1-2\lambda_{+}}{2},-\sqrt{2}kl\right)=0.\nonumber\\
&&
\label{spectr1}
\end{eqnarray}
A similar condition is derived at $K_{-}$ valley,
\begin{eqnarray}
&&\hspace{-10mm}
\lambda_{-}
U\left(\frac{1-2\lambda_{-}}{2},\sqrt{2}(k_0-k)l\right)
V\left(-\frac{1+2\lambda_{-}}{2},-\sqrt{2}kl\right)
-U\left(-\frac{1+2\lambda_{-}}{2},-\sqrt{2}kl\right)
V\left(\frac{1-2\lambda_{-}}{2},\sqrt{2}(k_0-k)l\right)=0.\nonumber\\
&&
\label{spectr2}
\end{eqnarray}
By solving Eqs.~(\ref{spectr1}) and (\ref{spectr2}) numerically, we determine
the dependence of dimensionless energy parameters $\lambda_{+}$ and
$\lambda_{-}$ on the wavevector $k$. The results for two different widths of
graphene ribbons, $W=5l$ and $W=10l$,  are shown in Fig.~\ref{fig.2}.

When the width of the ribbon is less than about 3 or 4 times the magnetic
length {$l \simeq 257 \mbox{\AA} /\sqrt{B [\mbox{T}]}$,}
we find that the spectra have little overlap with the usual bulk
spectra, i.e., $\lambda_{\pm}^{(bulk)}=n$ where $n$ is a
non-negative integer. Additionally, the separation between the nearest levels
quickly increases with decreasing $W$.

For the case $W=5l$, shown in the left panel of Fig.~\ref{fig.2}, only
the lowest level may have a hint at the middle plateau developing. However,
when the ribbon's width is larger than about 6 or 7 times the magnetic
length, nearly flat plateaus are already distinguishable  in the lowest levels
around the central wavevector $\frac12 k_0$. We also find that the lower the
level, the wider the plateau formed.

Let us also emphasize the following special feature of
the spectrum in a graphene ribbon with zigzag boundaries. As we
see from Fig.~\ref{fig.2}, for $\lambda_{+}\simeq 0$ (actually,
$E\simeq -\mu^{(+)}+\Delta^{(-)}$) and for $\lambda_{-}\simeq 0$
(actually, $E\simeq-\mu^{(-)}+\Delta^{(+)}$),
{dispersionless
surface solutions \cite{Brey,Abanin} exist at both valleys
(note that such solutions exist also in the case with no
magnetic field.\cite{Fujita,Brey1})}.
These solutions are bound to the $k\simeq 0$ and $k\simeq k_0$ edges
for the $K_{+}$ and $K_{-}$ valleys, respectively.
{It is noticeable that
unlike the case of a half-plane,\cite{GMSS2008} they cease to be dispersionless at
the opposite edges, i.e., at $k\simeq k_0$ ($k\simeq 0$) for the $K_{+}$
($K_{-}$) valley, respectively.}

Now, by restoring the spin index, we assemble the complete spectrum
of a graphene ribbon described by Hamiltonian (\ref{Hamiltonian})
with dynamical order parameters proposed in Refs.~\onlinecite{GGM2007,GGMS2008},
\begin{eqnarray}
\label{EsKpm}
E_{sK_{+}}^{(\pm)}(n,k) = -\mu_{s}^{(+)}\pm\sqrt{\lambda_{+}(n,k)\epsilon_0^2
+\left(\Delta_{s}^{(-)}\right)^2},&&\\
\label{EsKprime}
E_{sK_{-}}^{(\pm)}(n,k) = -\mu_{s}^{(-)}\pm\sqrt{\lambda_{-}(n,k)\epsilon_0^2
+\left(\Delta_{s}^{(+)}\right)^2}.&&
\end{eqnarray}
Notice that there exist eight sublevels that correspond to the lowest Landau
level. {Only half of these correspond to the {{\it bulk}} states, i.e., those
which remain normalizable on an infinite graphene plane.
The other half have wave functions localized only at
the edges.
In fact, using the properties of the parabolic cylinder functions,
one can show that in the coordinate space, the wave functions of
the additional branches of solutions (i.e., those which disappear on an infinite
plane) are localized near either $y=0$ or $y=W$ edges of the ribbon for {\it all}
values of the quantum number $k$ {(the {\it non-bulk} states).}
{\it In other words, the earlier mentioned
position-wavevector duality can be used only for the description of
the branches of the bulk states.}}

\begin{figure*}
\begin{center}
\includegraphics[width=.45\textwidth]{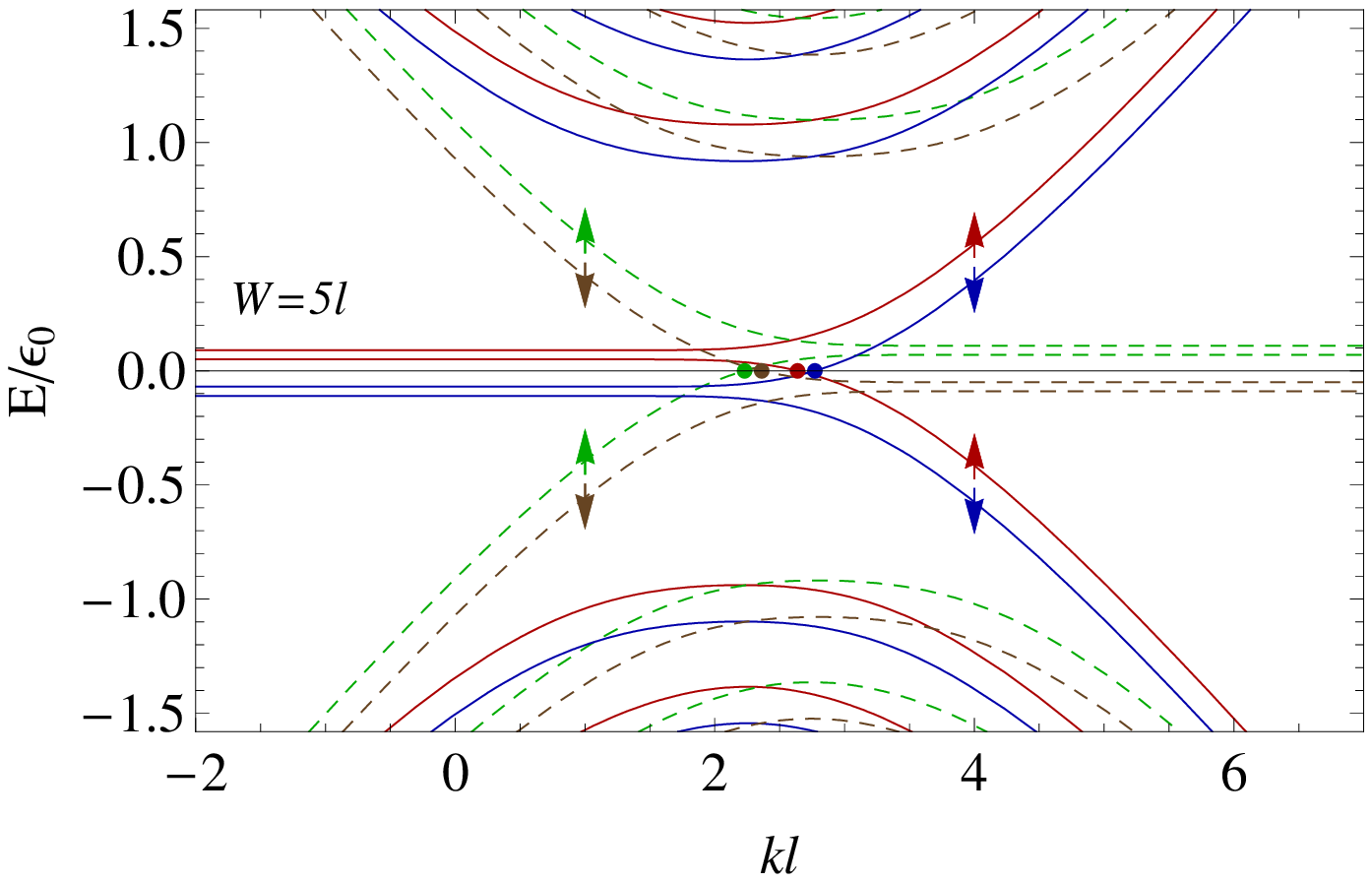}\hspace{.05\textwidth}
\includegraphics[width=.45\textwidth]{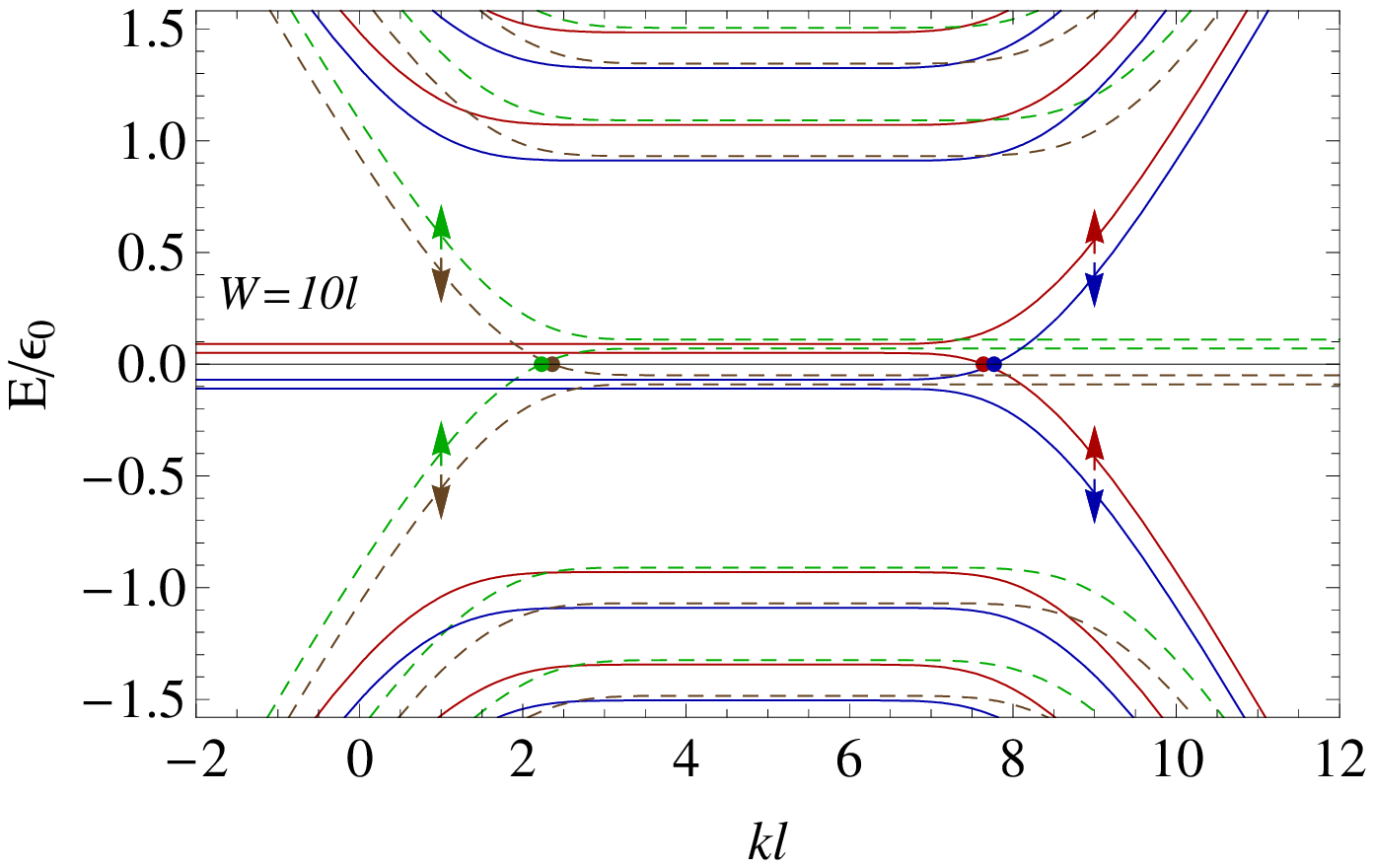}
\caption{(Color online) Numerical results for the low-energy spectra for ribbons with zigzag
edges. The ribbons' widths are $W=5l$ (left panel) and $W=10l$ (right panel).
The ferromagnetic gaps and dynamical masses  are as follows:
$\mu_{\pm}=\mp 0.08 \epsilon_0$, $\tilde\mu_{\pm}=0.01
\epsilon_0$, $\Delta_{\pm}=\pm 0.02 \epsilon_0$,
$\tilde\Delta_{\pm}=0$. The ferromagnetic gap dominates over the
mass gap, insuring the presence of gapless edge states (marked by
dots). The electron spins of the lowest energy sublevels are
marked by arrows. The spectra around the $K_{+}$ ($K_{-}$) point
are shown by solid (dashed) lines.}
\label{fig.3}
\end{center}
\end{figure*}

\begin{figure*}
\begin{center}
\includegraphics[width=.45\textwidth]{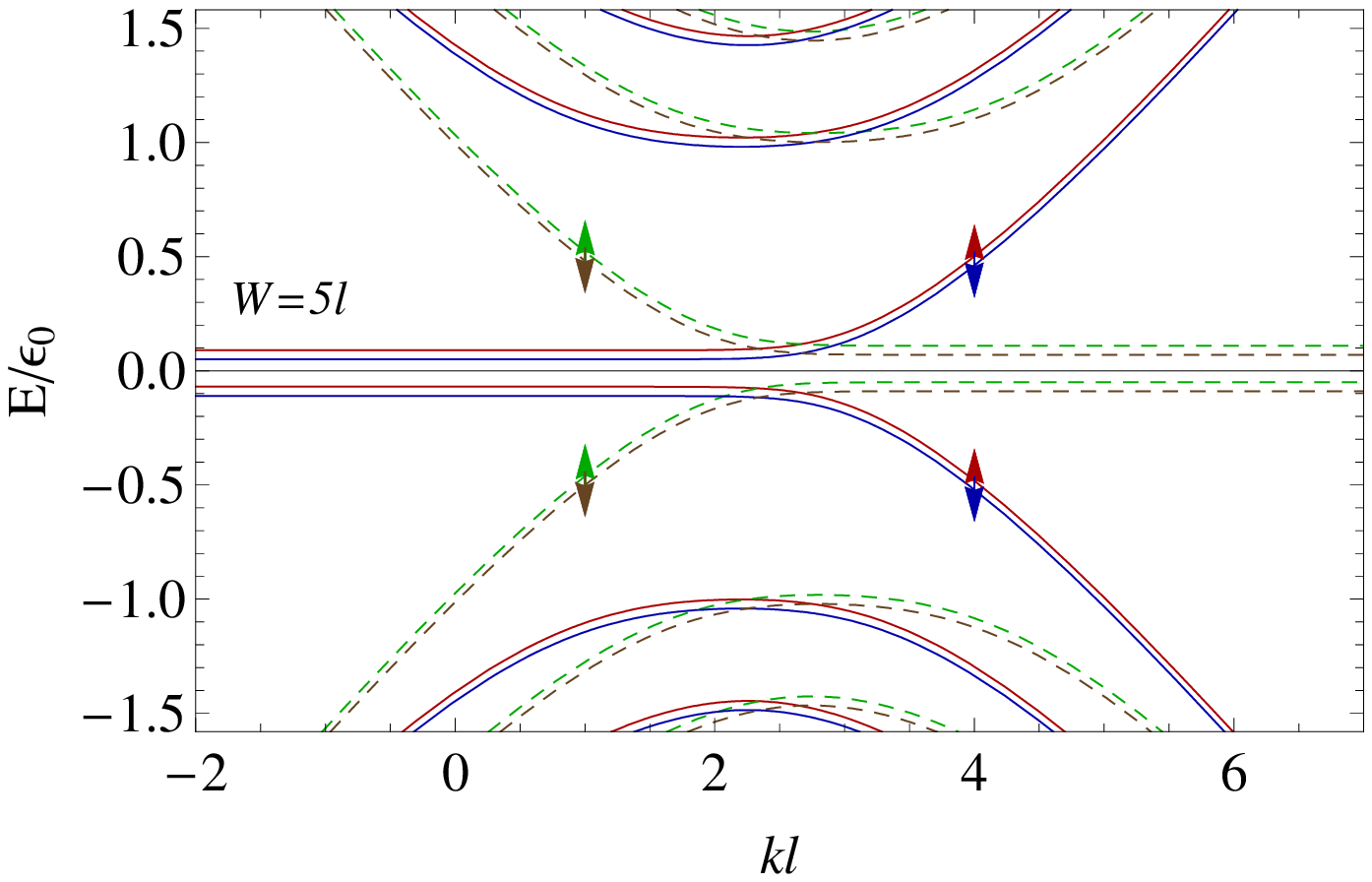}\hspace{.05\textwidth}
\includegraphics[width=.45\textwidth]{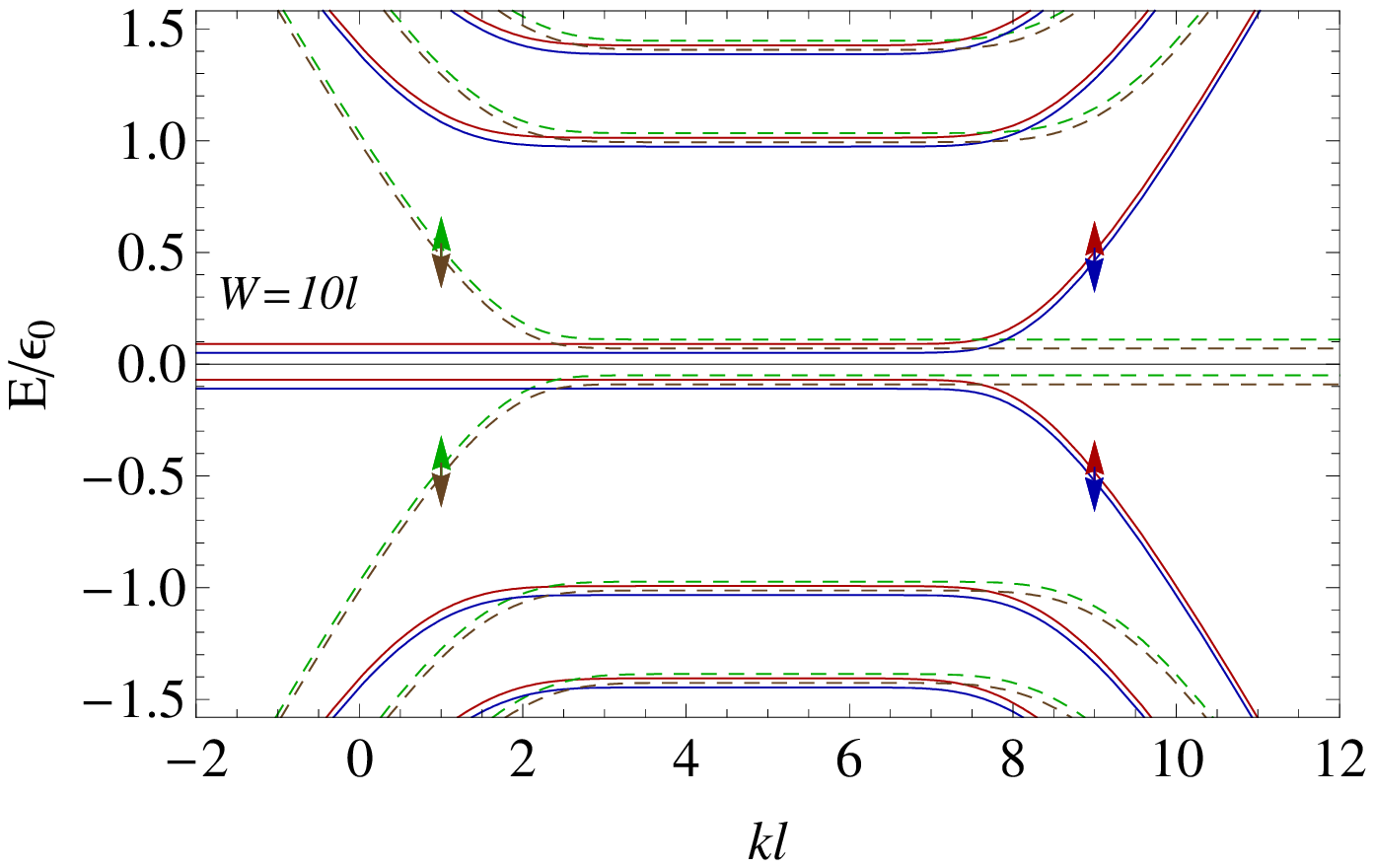}
\caption{(Color online) Same as Fig.~\ref{fig.3} but with the ferromagnetic gaps and dynamical masses
being as follows: $\mu_{\pm}=\mp 0.02 \epsilon_0$, $\tilde\mu_{\pm}=0.01 \epsilon_0$,
$\Delta_{\pm}=\pm 0.08 \epsilon_0$, $\tilde\Delta_{\pm}=0$. The mass gap dominates
over the ferromagnetic gap, insuring the absence of gapless edge states.}
\label{fig.4}
\end{center}
\end{figure*}

By making use of the numerical results for $\lambda_{+}$ and $\lambda_{-}$,
we can plot the actual energy spectra in the system. For the ribbons of widths
$W=5l$ and $W=10l$, these are presented in Figs.~\ref{fig.3} and \ref{fig.4}.
Here the choice of the order parameters resembles those of the solution
in Eq.~(\ref{singlet}) corresponding to the Dirac neutral point, i.e., the
$\nu=0$ plateau.\cite{GGM2007,GGMS2008} However, in order to lift the degeneracy
of all sublevels, we also added small non-zero values {for
the triplet chemical potentials $\tilde\mu_{\pm}$.}

By considering different relative strengths of ferromagnetic and mass
gaps in Figs.~\ref{fig.3} and \ref{fig.4}, we see that there exist gapless
edge states (whose energy vanishes at certain values of $k$) only when
the ferromagnetic gap dominates over the mass gap, i.e., $|\mu_s^{(\pm)}|
>|\Delta_s^{(\mp)}|$. (The values of wavevectors that give gapless modes
are marked by the dots in the spectra in Fig.~\ref{fig.3}.) From
Fig.~\ref{fig.2}, we can see that $\lambda_{\pm}(0,k)$ is nonnegative
and approaches zero at certain values of the wavevector. This feature
together with dispersion relations (\ref{EsKpm}) and (\ref{EsKprime})
makes it clear that the necessary and sufficient condition for the
existence of gapless modes is that at least one of the inequalities
$|\mu_s^{(-)}|>|\Delta_s^{(+)}|$, $|\mu_s^{(+)}|>|\Delta_s^{(-)}|$
is satisfied for at least one spin choice. 

So far, we have considered only the case with nonzero {\em singlet} Dirac masses.
However, as is clear from Eqs.~(\ref{EsKpm}) and (\ref{EsKprime}),
the results for nonzero {\em triplet} Dirac masses will look exactly
the same as in Figs.~\ref{fig.3} and \ref{fig.4}.
At first sight, this might
appear to be surprising because the symmetry properties of the two types of
the masses are different: while the triplet masses break the $SU(2)_s$ valley
symmetries, the singlet masses do not. Let us turn to the discussion of this
point.
First of all, the corresponding symmetry is exact only on an infinite plane. As
to a finite width ribbon, it is explicitly broken by the boundary conditions there,
as seen from the comparison of Eq.~(\ref{spectr1}) for $K_{+}$ valley and
Eq.~(\ref{spectr2}) for
$K_{-}$ valley. This is also obvious from the solutions for $\lambda_{\pm}$
in Fig.~\ref{fig.2}.
On the other hand, although one might expect that
the symmetry arguments are to be approximately
applicable to the states with intermediate values of the wavevector $k$
in a bulk of a ribbon, both these types of the Dirac masses still lead to
the same spectra there. This puzzle is resolved as follows. As was already
emphasized above, one should distinguish between the bulk
states and non-bulk ones on a ribbon. The dispersion relations in
the ribbon bulk for the former approximately coincide with those of the states
on an infinite plane. In the case of the LLL, they are
$E_{sK_{+}}=-\mu_s^{(+)}+\Delta_s^{(-)}$
and $E_{sK_{-}}=-\mu_s^{(-)}+\Delta_s^{(+)}$.\cite{GGM2007,GGMS2008}
{The point is that} on the ribbon, the asymmetry of this spectrum
for $\tilde{\Delta}_s \neq 0$
is washed away by the inclusion of four additional
{\it non-bulk} branches, which are localized
exclusively on the edges and whose energies are approximately given
by similar expressions but with the opposite signs in front of $\Delta_s^{(\pm)}$.
As for higher LLs, it is known that their energies do not depend on the type
of Dirac masses even on an infinite plane, provided only a mass of one type
is present.\cite{GGM2007,GGMS2008}
{\it Thus, on a ribbon, the sublevel structure of the LLL becomes similar to that
of higher LLs.}

It is instructive to compare the properties of the solutions on a finite
width ribbon with those on a half-plane.\cite{GMSS2008} First of all,
instead of the eight sublevels of the LLL on a ribbon, there are only six
ones on a half-plane. The reason of that is of course connected with
the fact that each edge adds two additional sublevels connected with
two orientations of spin. Secondly, unlike the case of a ribbon,
there are LLL sublevels which are dispersionless for
{\it all} values of $k$ on a half-plane. This feature can be also easily
understood. As the width $W$ of a ribbon goes to infinity, the edge $k = k_0$
disappears and the solutions which are bound to the edge $k = 0$
become dispersionless for all values $k > 0$.

In relation to the dispersionless modes, we would like to add the
following 
{comment.} Because of level crossing, there is an ambiguity in the
definition of a single branch of solutions in the LLL. By mixing together
the two branches associated with different valleys, one can construct a
new branch that is approximately dispersionless for all values of $k$, just
as seen in numerical calculations.\cite{Arikawa} As should be clear from
our analytical analysis, however, such a branch is only approximately
dispersionless for intermediate values of the wavevector $k$.

\section{Numerical results in the case of armchair edges.}
\label{armchair-num}

Let us now consider the case of the armchair edges.
Since the armchair boundary conditions (\ref{BC-arm1}) and (\ref{BC-arm2})
mix the chiralities associated with the $K_+$ and $K_-$ valleys, this case
is essentially more complicated for both the analytic and numerical analyses
on a ribbon.

The explicit form
of the boundary conditions (\ref{BC-arm1}) and (\ref{BC-arm2})
reads:
\begin{eqnarray}
C_{1} \frac{E+\mu^{(+)}-\Delta^{(-)}}{\epsilon_0 }
U\left(\frac{1-2\lambda_{+}}{2},\sqrt{2}{kl}\right)
+C_{2} V\left(\frac{1-2\lambda_{+}}{2},\sqrt{2}{kl}\right)+\nonumber\\
+ C_{3}U\left(-\frac{1+2\lambda_{-}}{2},\sqrt{2}{kl}\right)
+C_{4} \frac{E+\mu^{(-)}+\Delta^{(+)}}{\epsilon_0}
V\left(-\frac{1+2\lambda_{-}}{2},\sqrt{2}{kl}\right)&=&0,
\label{cond1arm}\\
C_{1} U\left(-\frac{1+2\lambda_{+}}{2},\sqrt{2}{kl}\right)
+C_{2}\frac{E+\mu^{(+)}+\Delta^{(-)}}{\epsilon_0 }
V\left(-\frac{1+2\lambda_{+}}{2},\sqrt{2}{kl}\right)+\nonumber\\
+C_{3} \frac{E+\mu^{(-)}-\Delta^{(+)}}{\epsilon_0}
U\left(\frac{1-2\lambda_{-}}{2},\sqrt{2}{kl}\right)
+C_{4}V\left(\frac{1-2\lambda_{-}}{2},\sqrt{2}{kl}\right)&=&0,
\label{cond2arm}\\
C_{1} \frac{E+\mu^{(+)}-\Delta^{(-)}}{\epsilon_0 }
U\left(\frac{1-2\lambda_{+}}{2},\sqrt{2}{(k-k_0)l}\right)
+C_{2} V\left(\frac{1-2\lambda_{+}}{2},\sqrt{2}{(k-k_0)l}\right)+\nonumber\\
+ C_{3}U\left(-\frac{1+2\lambda_{-}}{2},\sqrt{2}{(k-k_0)l}\right)
+C_{4} \frac{E+\mu^{(-)}+\Delta^{(+)}}{\epsilon_0}
V\left(-\frac{1+2\lambda_{-}}{2},\sqrt{2}{(k-k_0)l}\right)&=&0,
\label{cond3arm}\\
C_{1} U\left(-\frac{1+2\lambda_{+}}{2},\sqrt{2}{(k-k_0)l}\right)
+C_{2}\frac{E+\mu^{(+)}+\Delta^{(-)}}{\epsilon_0 }
V\left(-\frac{1+2\lambda_{+}}{2},\sqrt{2}{(k-k_0)l}\right)+\nonumber\\
+C_{3} \frac{E+\mu^{(-)}-\Delta^{(+)}}{\epsilon_0}
U\left(\frac{1-2\lambda_{-}}{2},\sqrt{2}{(k-k_0)l}\right)
+C_{4}V\left(\frac{1-2\lambda_{-}}{2},\sqrt{2}{(k-k_0)l}\right)&=&0,
\label{cond4arm}
\end{eqnarray}
where $k_0=-W/l^2$.
It is a homogeneous system of four linear equations for four unknown constants $C_i$, $i=1,2,3,4$.
The nontrivial solution exists when the determinant of a matrix made of the coefficients is
zero. We numerically solve the corresponding equation to determine the spectrum of single-particle
states in the graphene ribbon.

\begin{figure}
\begin{center}
\includegraphics[width=.45\textwidth]{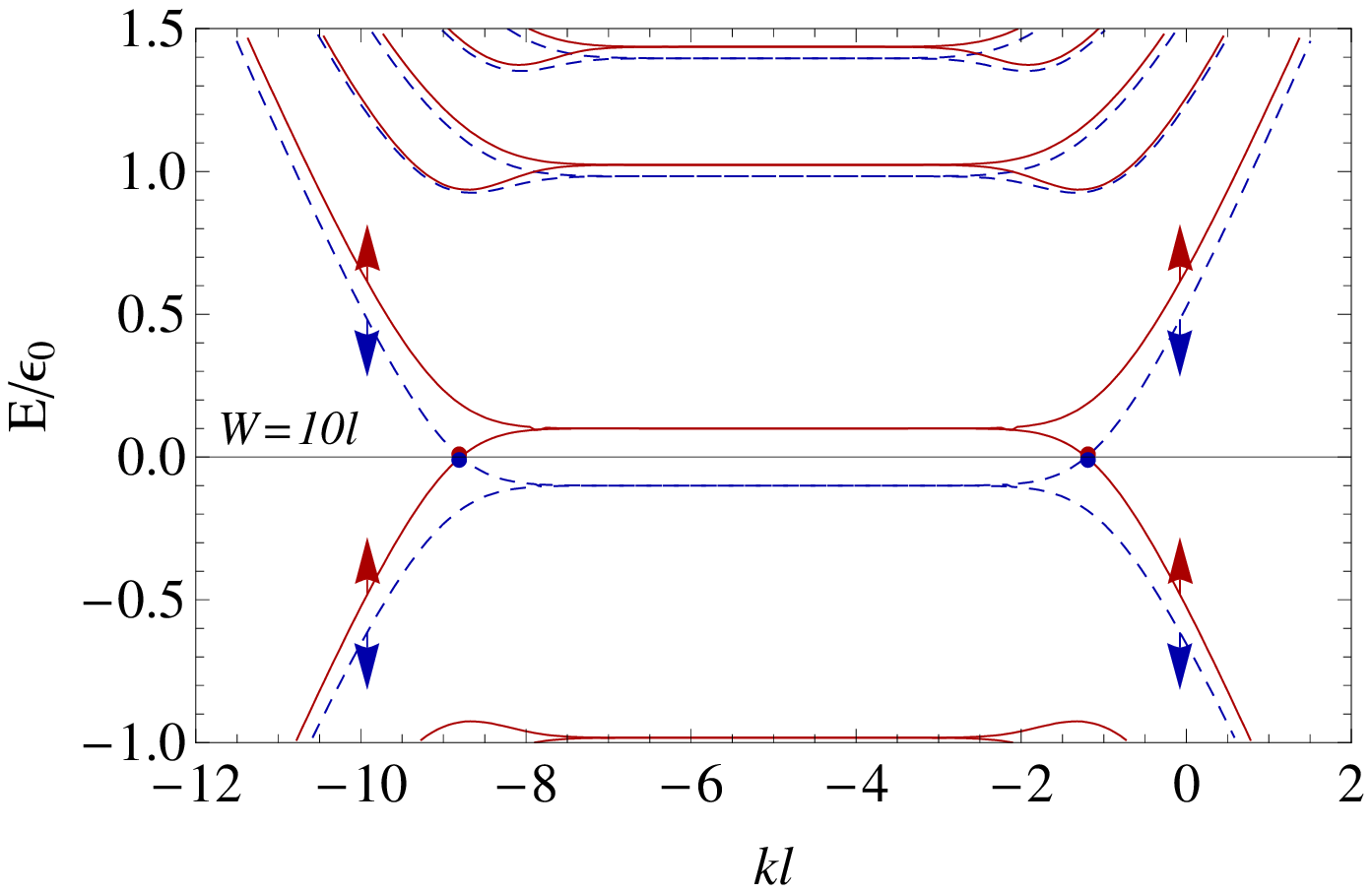}\hspace{.05\textwidth}
\includegraphics[width=.45\textwidth]{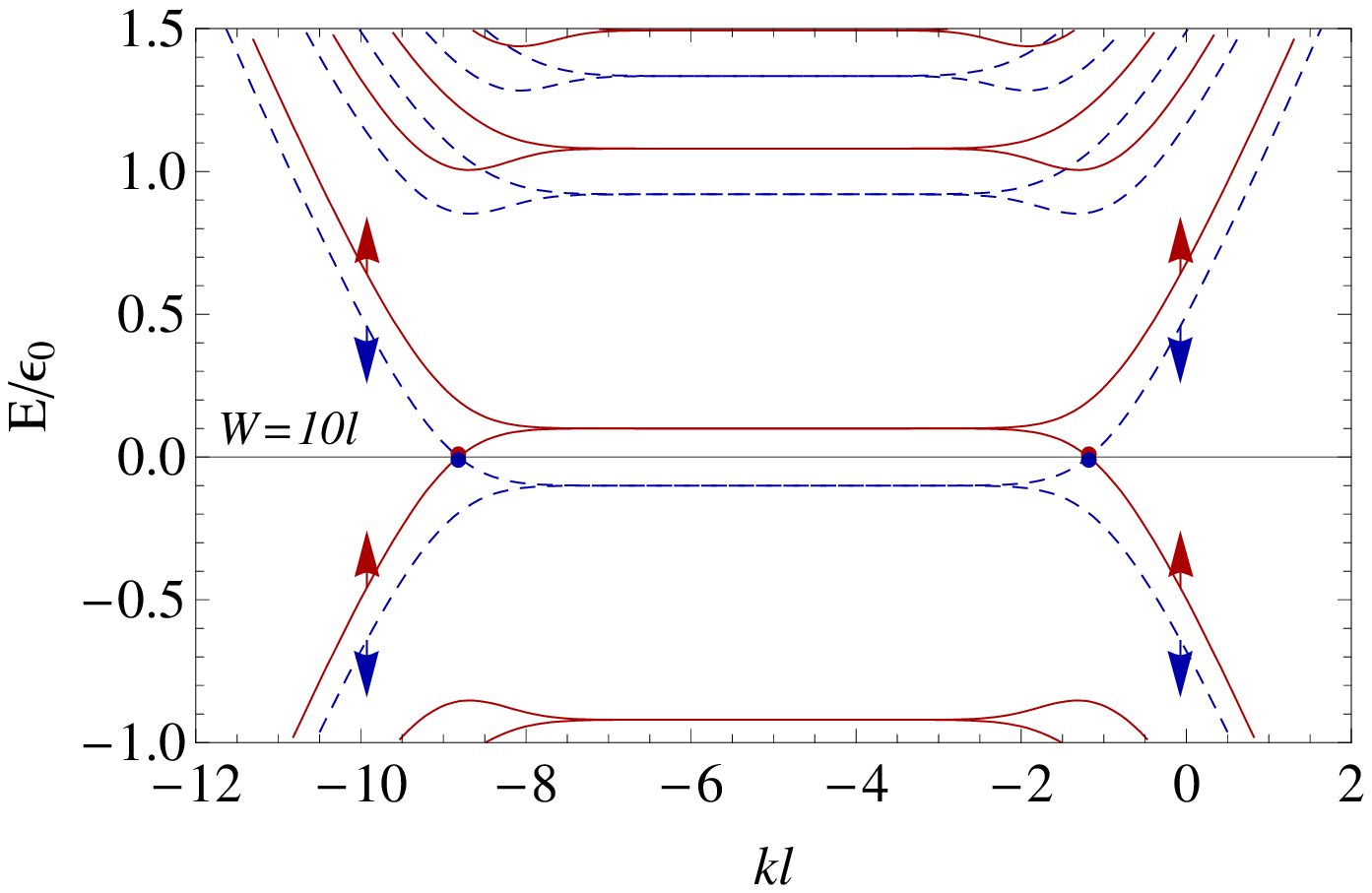}
\caption{(Color online)
Numerical results for the low-energy spectra for a ribbon with armchair edges of width
$W = 10l$ in the case of nonzero spin splitting and nonzero {\em singlet} masses.
Nonzero dynamical parameters are as follows: $\mu_{\pm}= \mp 0.02\epsilon_0$,
$\Delta_{\pm}= \pm 0.08\epsilon_0$ (left panel) and $\mu_{\pm}= \mp 0.08\epsilon_0$,
$\Delta_{\pm}= \pm 0.02\epsilon_0$ (right panel). Gapless edge states (marked by
dots) are present in both cases. The electron spins of the lowest energy sublevels
are marked by arrows.}
\label{fig.arm.siglet}
\end{center}
\end{figure}

The cases with nonzero singlet ($\Delta_{s}$) and triplet ($\tilde\Delta_{s}$)
Dirac masses are studied separately. Numerical results for singlet masses are shown in
Fig.~\ref{fig.arm.siglet}. We find that gapless edge states always appear when
there are only Zeeman-like (QHF) gap and singlet Dirac masses $\Delta_{s}$,
irrespective of the actual relation between their values.
{We conclude, therefore, that for the case of a singlet Dirac mass,
the condition for the existence of gapless modes is less
constrained at the armchair edges than at the zigzag ones. In fact, this result
should have been expected after recalling that the singlet Dirac masses do not
break the global valley symmetry groups $SU(2)_s\subset U(2)_s$. These symmetries
(one for spin up and the other for spin down) protect the double degeneracy of the
Landau levels in the bulk. The property of such sublevels in the LLL is that they
repel in opposite directions near the edges both in the absence of
singlet Dirac masses and when they are present. The latter leads to gapless states.
(Note that in this argument we implicitly use the fact
that the energy separation of the sublevels may become arbitrarily large. This is
not true on a lattice. However, even on the lattice the energy separation may
become much larger than the dynamical scale of the mass, suggesting that the
conclusion is still valid in that case.) In fact, the absence of gapless modes on a ribbon
with zigzag edges {and $|\Delta_s|> |\mu_s|$}
is a special property, which is related to the dispersionless nature
of the LLL modes at the edges with $k \simeq 0$ or $k \simeq k_0$.}

\begin{figure}
\begin{center}
\includegraphics[width=.48\textwidth]{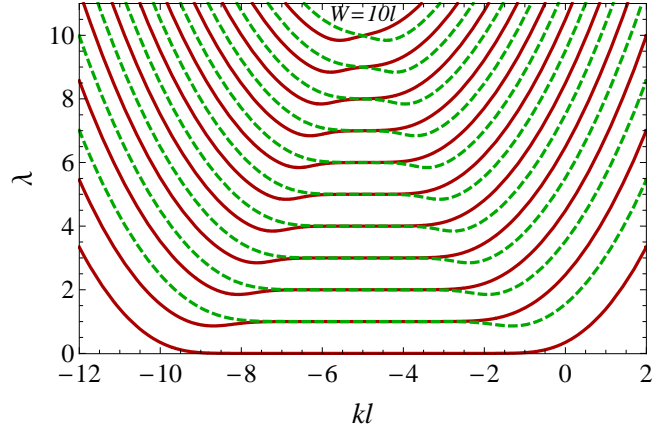}
\caption{(Color online) Numerical solutions of Eq.~(\ref{equation:arm-triplet}) for the dimensionless
parameter $\lambda=[(E+\mu)^2-\tilde{\Delta}^2]/\epsilon_0^2$ in the case of armchair
edges when $\tilde{\mu}=\Delta=0$. In the higher Landau levels, $n\geq 1$, the solutions
to the equations $f_{-}(k,\lambda)=0$ and $f_{+}(k,\lambda)=0$ are shown by solid and
dashed lines, respectively.}
\label{lambda-arm}
\end{center}
\end{figure}

{Let us now turn to a triplet Dirac mass.
We found that the analysis of the energy spectra can be considerably
simplified in the case of nonzero triplet Dirac masses but vanishing
$\tilde{\mu}_s$ and $\Delta_s$. The central observation is that in
this case the determinant of the matrix corresponding to
Eqs.~(\ref{cond1arm})--(\ref{cond4arm}) can be reduced to
factorized form, and the spectral equation becomes
\begin{equation}
f_{+}(k,\lambda) f_{-}(k,\lambda) =0,
\label{equation:arm-triplet}
\end{equation}
where, by definition, $\lambda=[(E+\mu)^2-\tilde{\Delta}^2]/\epsilon_0^2$ and
\begin{eqnarray}
f_{\pm}(k,\lambda) &=&
 U\left(-\frac{1+2\lambda}{2},\sqrt{2}{(k-k_0)l}\right)
 V\left( \frac{1-2\lambda}{2},\sqrt{2}{kl}\right)
-U\left(-\frac{1+2\lambda}{2},\sqrt{2}{kl}\right)
 V\left( \frac{1-2\lambda}{2},\sqrt{2}{(k-k_0)l}\right)\nonumber\\
&+&\lambda\left[
 U\left( \frac{1-2\lambda}{2},\sqrt{2}{(k-k_0)l}\right)
 V\left(-\frac{1+2\lambda}{2},\sqrt{2}{kl}\right)
-U\left( \frac{1-2\lambda}{2},\sqrt{2}{kl}\right)
 V\left(-\frac{1+2\lambda}{2},\sqrt{2}{(k-k_0)l}\right)
\right]\nonumber\\
&\pm & \sqrt{\lambda}\left[
 U\left( \frac{1-2\lambda}{2},\sqrt{2}{(k-k_0)l}\right)
 V\left( \frac{1-2\lambda}{2},\sqrt{2}{kl}\right)
-U\left( \frac{1-2\lambda}{2},\sqrt{2}{kl}\right)
 V\left( \frac{1-2\lambda}{2},\sqrt{2}{(k-k_0)l}\right)\right.
\nonumber\\
&+&
\left.
 U\left(-\frac{1+2\lambda}{2},\sqrt{2}{(k-k_0)l}\right)
 V\left(-\frac{1+2\lambda}{2},\sqrt{2}{kl}\right)
-U\left(-\frac{1+2\lambda}{2},\sqrt{2}{kl}\right)
 V\left(-\frac{1+2\lambda}{2},\sqrt{2}{(k-k_0)l}\right)
\right].\nonumber\\
\end{eqnarray}
Restoring the spin index again, the complete energy spectrum takes the form:
\begin{equation}
E_{sn}(k) = -\mu_{s} \pm \sqrt{ \lambda (kl, n)\epsilon_0^2 +\tilde{\Delta}_{s}^2},
\label{spectr_arm-triplet}
\end{equation}
where $n=0,1,2,\ldots$. Our numerical results for several lowest branches
of $\lambda$ versus $kl$ are presented in Fig.~\ref{lambda-arm}. Note that
$\lambda(kl, n)\simeq n$ at intermediate values of $kl$. This means that
the low-energy spectrum of a finite width ribbon far from the edges should
be approximately the same as the spectrum of graphene on an infinite plane.
}

\begin{figure*}
\begin{center}
\includegraphics[width=.45\textwidth]{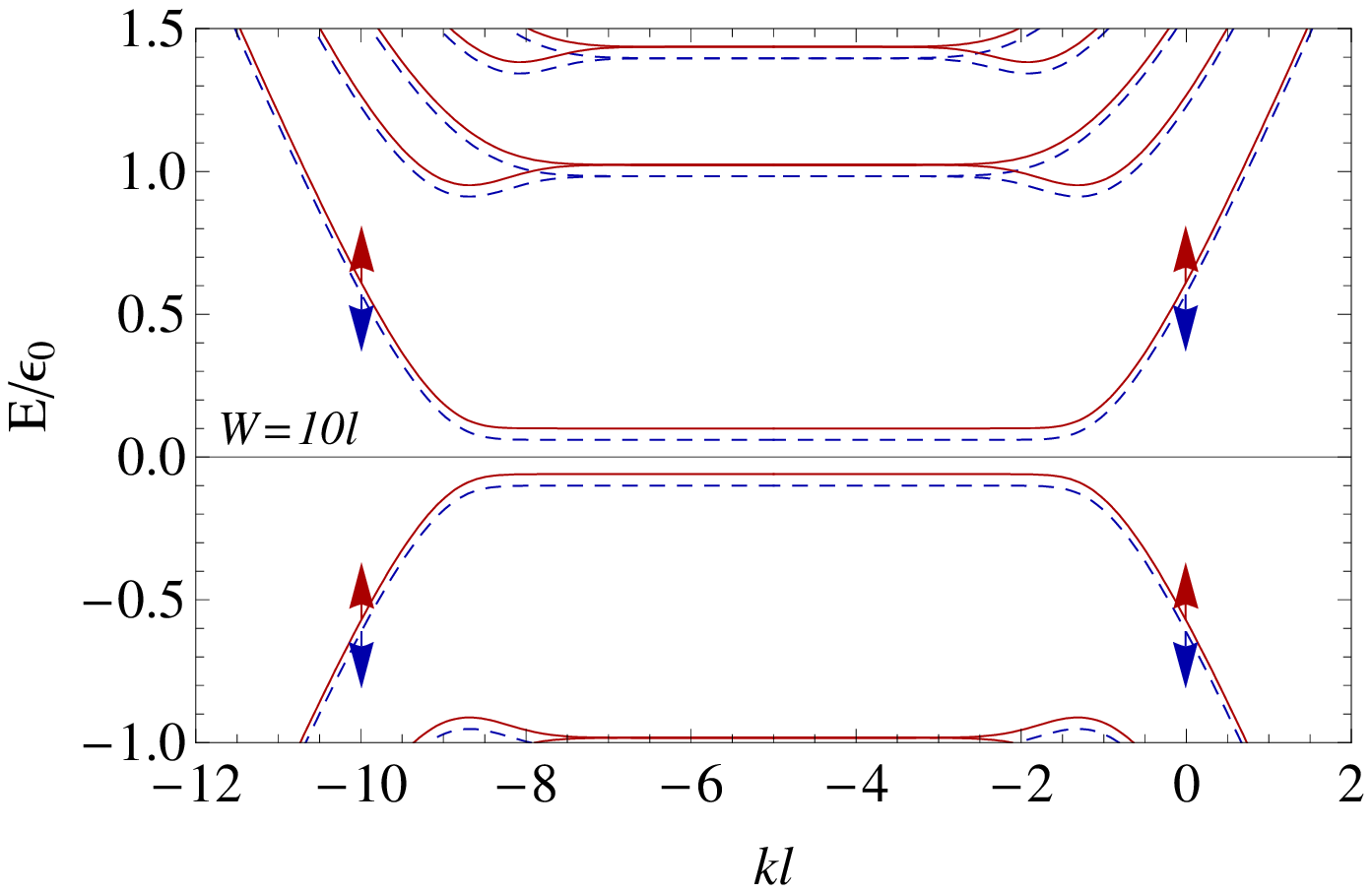}\hspace{.05\textwidth}
\includegraphics[width=.45\textwidth]{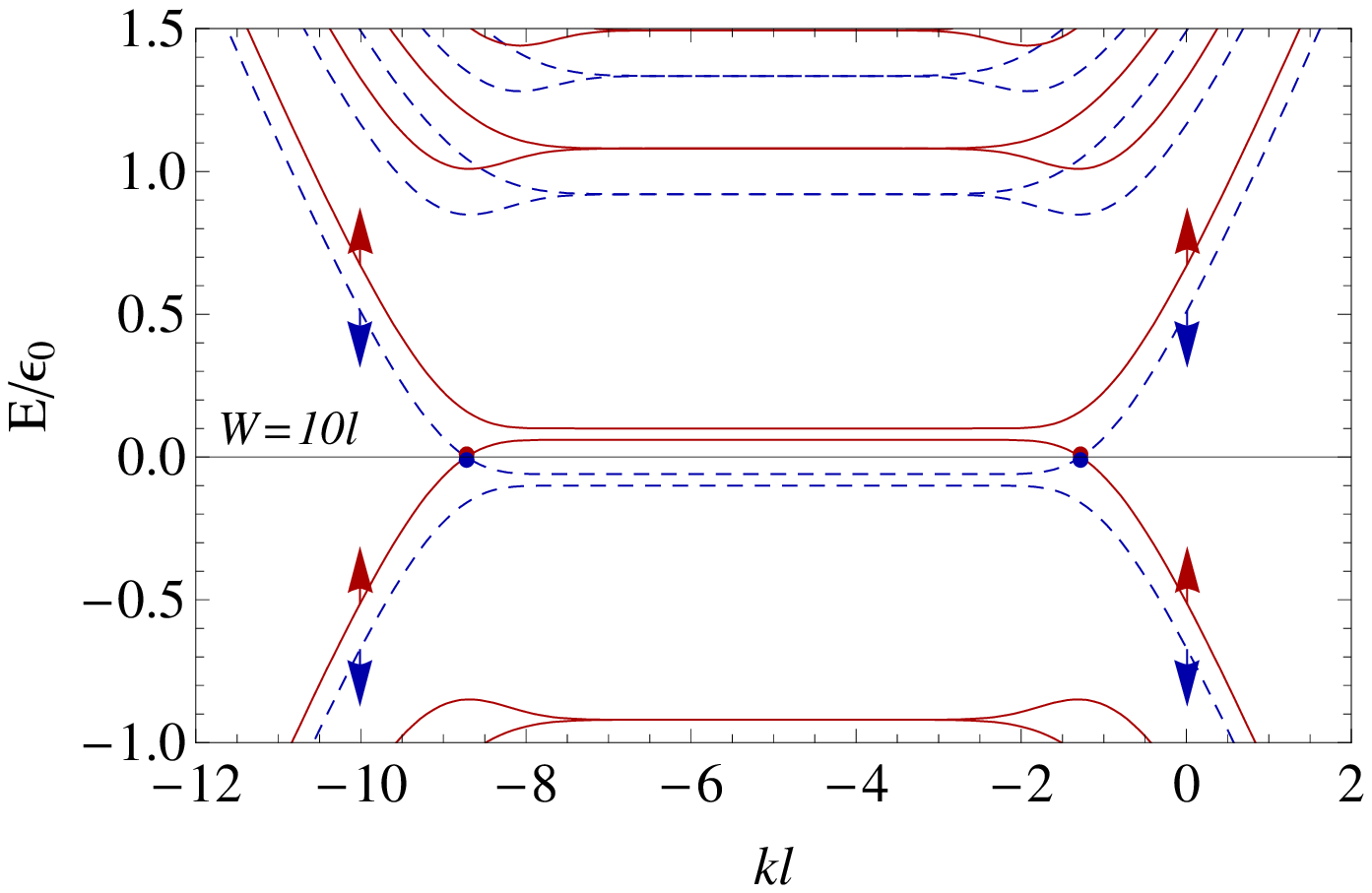}
\caption{(Color online) Same as Fig.~\ref{fig.arm.siglet}, but for the case of nonzero {\em triplet} masses.
Nonzero dynamical parameters are as follows: $\mu_{\pm}= \mp 0.02\epsilon_0$,
{$\tilde\Delta_{\pm}= 0.08\epsilon_0$} (left panel) and $\mu_{\pm}= \mp 0.08\epsilon_0$,
{$\tilde\Delta_{\pm}= 0.02\epsilon_0$} (right panel). The existence of gapless modes
(marked by dots) depends on the relative magnitude of $|\mu_{\pm}|$ and $|\tilde \Delta_{\pm}|$.
The electron spins of the lowest energy sublevels are marked by arrows.}
\label{fig.arm.triplet}
\end{center}
\end{figure*}

{
The energy spectrum in the case of triplet Dirac masses are shown in
Fig.~\ref{fig.arm.triplet}. As seen from the figure, the presence of gapless modes in the
spectrum sensitively depends on the relation between the values of the Zeeman-like (QHF)
gaps and the triplet Dirac masses $\tilde\Delta_{s}$. Such modes exist only when the
magnitude of the Dirac mass is less than the QHF gap. The physics underlying
this result is clear. When $|\tilde\Delta_{s}|> |\mu_{s}|$, the valley splitting is large
and there are two LLL branches of states with opposite signs of their energies for each
direction of the spin [see Eq.~(\ref{spectr_arm-triplet}) and the left panel in
Fig.~\ref{fig.arm.triplet}]. As a result, repelling of these branches at the edges
does not lead to the creation of gapless modes. On the other hand, in the case of
$|\tilde\Delta_{s}|< |\mu_{s}|$, shown in the right panel of Fig.~\ref{fig.arm.triplet}
[see also Eq.~(\ref{spectr_arm-triplet})], the valley splitting is small, the energies
of the two branches have the same sign, and the process of their repelling at the edges
inevitably creates gapless modes. In essence, this is the same condition as for zigzag
edges in a system with triplet Dirac masses studied in the previous section,}
\begin{equation}
|\tilde\Delta_{s}|< |\mu_{s}|.
\label{triplet}
\end{equation}

\section{Discussion}
\label{Discussion}

The main result in this paper is establishing the criterion for the existence of
gapless modes among the edge states in a graphene ribbon with zigzag and armchair
type edges. The method used in this paper combines analytic and
numerical approaches that allows to get a deeper insight into the nature of
edge states.

In the case of zigzag edges, gapless modes exist when the ferromagnetic
(Zeeman-like) gap dominates over the mass gap of any type, or, more formally,
when any of the conditions $|\mu_s^{(\pm)}|>|\Delta_s^{(\mp)}|$ are satisfied
for at least one spin choice $s=\pm$. This is consistent with the two limiting
cases analyzed in Ref.~\onlinecite{AbaninNovoselov}.

For a ribbon with armchair edges, the condition for the existence of gapless
modes is more involved however. In this case, it depends on the actual type (singlet or
triplet) of the dynamical Dirac mass induced. For singlet Dirac masses, there are
always exist gapless modes. On the other hand, for triplet masses,
gapless modes exist only when there is a sufficiently large
ferromagnetic (Zeeman-like) gap that dominates over the masses.

One of the most interesting consequences of our finding here is a possibility of
resolving the seemingly contradicting interpretations of the $\nu=0$ plateau in
terms of either quantum Hall metal or insulator regimes.\cite{AbaninNovoselov,Ong2007}
As follows from Eq.~(\ref{singlet}), the criterion for the existence of gapless
edge states at the Dirac neutral point takes the simple form $Z + A > M$ {(note
that since a boundary of real graphene samples consists of both zigzag and
armchair edges,\cite{review1}
it is appropriate to use the more constrained condition for the existence of gapless
modes corresponding to the zigzag edges).}
As was pointed in Refs.~\onlinecite{GGM2007,GGMS2008},
this condition implies the existence of a critical value of the transverse magnetic
field $B^{(cr)}$, where the insulator regime switches to the metallic one.
A crude estimate for this value yields
$0.01\, {\rm T} \lesssim B^{(cr)} \lesssim 200\,{\rm T}$
for an effective coupling constant $\lambda$ taken in the interval
$0.02\, \lesssim \lambda \lesssim 0.2$.\cite{GGMS2008}
As one can see, $B^{(cr)}$ is very sensitive to the choice of
$\lambda$. In order to fix the values of $\lambda$ and $B^{(cr)}$
more accurately, one should utilize more realistic models of graphene
that incorporate consistently disorder among other things.\cite{disorder1,disorder2}
This is a topic for future studies however.

There is another possibility to explain the experimental results in 
Ref.~\onlinecite{Ong2007}. As was shown in Ref.~\onlinecite{GGMS2008},
besides the $S1$ solution (\ref{singlet}), there is another, triplet ($T$),
solution around the Dirac neutral point corresponding to the $\nu = 0$ plateau,
in the model of Refs.~\onlinecite{GGM2007,GGMS2008}. In the $T$ solution, while
both spin up and spin down quasiparticle states have a triplet Dirac mass
$\tilde{\Delta}_\pm = M$, the chemical potentials $\mu_{\pm}$ are small, 
$\mu_{\pm} = \mp Z$. Therefore, there are no gapless edge states for this 
solution and it describes the quantum Hall insulator regime. Calculating the 
difference of the free energy densities for these two solutions, it was 
shown\cite{GGMS2008} that it is the Zeeman term which makes the $S1$ solution 
more favorable: without it, the $S1$ and $T$ solutions would correspond to 
two degenerate ground states. On the other hand, as was pointed out in 
Ref.~\onlinecite{qhf4} (see also Refs.~\onlinecite{dm2,Aleiner2007,Herbut0610}),
there are small on-site repulsion interaction terms on the graphene lattice which 
favor the triplet solution (such terms were ignored in the model in 
Refs.~\onlinecite{GGM2007,GGMS2008}). It would be interesting to figure out 
the role of these terms in choosing the genuine ground state in the
present dynamics {at different values of a magnetic field.}

{In the future, it would be interesting to extend the present analysis 
by considering inhomogeneous QHF and MC order parameters, which should be 
consistently determined from the gap equation. Such inhomogeneous order 
parameters on a ribbon with zigzag edges in a magnetic field can be expected 
because they exist on ribbons with the zigzag edges in the absence of a magnetic 
field.\cite{Fujita,Yamashiro2003PRB,Fernandez2008PRB,Jung2008} The point is that as 
a consequence of the nearly dispersionless (flat) subbands, a peak in the density 
of states occurs near the zigzag edge, resulting in an ordered magnetic phase 
even at zero magnetic field.\cite{Fujita} In the presence of a strong magnetic 
field, however, the MC and QHF order parameters alone remove the degeneracy 
of the dispersionless states with opposite spins (or pseudospins, related to 
the $SU(2)_s$ symmetries discussed in Sec.~\ref{secII}). This means 
that the edge magnetism (or pseudomagnetism) may at least partially be captured 
by the homogeneous order parameters already included in this paper. Of course, 
it would be  important to reexamine these arguments in more detail in future 
studies.}

\acknowledgments The authors acknowledge discussions with E.~Gorbar,
H.~Fertig, I.~Herbut, L.~Levitov, and B.~Shklovskii. The work of
V.P.G. was supported by the SCOPES-project IB 7320-110848 of the
Swiss NSF, the grant 10/07-N ``Nanostructure systems, nanomaterials,
nanotechnologies", and by the Program of Fundamental Research of the
Physics and Astronomy Division of the National Academy of Sciences
of Ukraine. V.A.M. and C.M.W. acknowledge the support of
the Natural Sciences and Engineering Research Council of Canada.
The work of S.G.S. was supported by the Program of Fundamental
Research of the Physics and Astronomy Division of the National
Academy of Sciences of Ukraine. The work of I.A.S. was supported by the
start-up funds from the School of Applied Arts and Sciences at the 
Arizona State University. This work was made possible by the facilities 
of the Shared Hierarchical Academic Research Computing
Network (SHARCNET:www.sharcnet.ca).

\end{document}